\documentclass[%
reprint,
 amsmath,amssymb,
 aip,
]{revtex4-2}

\usepackage{float}

\newcommand{\ket}[1]{\vert #1 \rangle}
\newcommand{\Ket}[1]{\big\vert #1 \big\rangle}
\newcommand{\bra}[1]{\langle #1 \vert}

\newcommand{\Dbraket}[2]{\langle #1 \hspace{.10em} \vert \hspace{.10em}  #2 \rangle}

\newcommand{\Tbraket}[3]{\langle #1 \hspace{.10em} \vert \hspace{.10em} #2 \hspace{.10em} \vert \hspace{.10em} #3 \rangle}

\newcommand{\ketR}{\ket{\mathrm{R}}}
\newcommand{\tbar}{\boldsymbol{\bar{t}}}
\newcommand{\simH}{\bar{H}}
\newcommand{\simP}{\bar{\mathcal{P}}}
\newcommand{\simS}{\bar{{S}}}
\renewcommand{\P}{\mathcal{P}}
\newcommand{\PP}{P^{\alpha}P^{\beta}}
\newcommand{\QQ}{Q^{\alpha}Q^{\beta}}
\newcommand{\PQa}{P^{\beta}Q^{\alpha}}
\newcommand{\PQb}{P^{\alpha}Q^{\beta}}
\newcommand{\braR}{\bra{\mathrm{R}}}
\newcommand{\sts}{\sin^2\theta}

\newcommand{\cts}{\cos^2\theta}
\newcommand{\brad}{\bra{^{ab}_{ij}}}

\newcommand{\brabj}{\bar{\bra{^{b}_{j}}}}
\newcommand{\Eb}[1]{E_{#1}^{\beta}}
\newcommand{\Ea}[1]{E_{#1}^{\alpha}}

\newcommand{\revHK}[1]{{\color{black}#1}}
\newcommand{\revS}[1]{\textcolor{black}{#1}}
\newcommand{\sarai}[1]{\textcolor{black}{#1}}

\usepackage{multirow}
\usepackage{siunitx}
\DeclareSIUnit\angstrom{\text{Å}}
\usepackage[utf8]{inputenc}
\usepackage{comment}
\usepackage{xcolor}
\usepackage{lipsum}
\usepackage{soul}

\usepackage{float}
\usepackage{graphicx}
\usepackage{dcolumn}
\usepackage{bm}

\usepackage{booktabs}

\begin{document}

\preprint{APS/123-QED}

\title{Entanglement Coupled Cluster Theory: Exact Spin-Adaptation}

\author{Sarai Dery Folkestad}
\email{sarai.d.folkestad@ntnu.no}
\affiliation{%
Department of Chemistry, Norwegian University of Science and Technology, 7491 Trondheim, Norway}%
\author{Bendik Støa Sannes}%
\affiliation{%
Department of Chemistry, Norwegian University of Science and Technology, 7491 Trondheim, Norway}%
\author{Henrik Koch}%
\email{henrik.koch@ntnu.no}
\affiliation{%
Department of Chemistry, Norwegian University of Science and Technology, 7491 Trondheim, Norway}%
\affiliation{Scuola Normale Superiore, Piazza dei Cavaleri 7, 56126 Pisa, Italy}

\date{\today}

\begin{abstract}
We present a novel \revHK{framework for} spin-adapted coupled cluster theory. \revHK{The} approach {exploits} the entanglement of an open-shell molecule with electrons in a non-interacting bath. \sarai{Together, the molecule and the bath} form a closed-shell \sarai{system, and} electron correlation can be \revHK{included using} the standard spin-adapted closed-shell coupled cluster formalism.
A projection operator, which enforces conditions on the electrons in the bath, is used to obtain the desired state of the \sarai{molecule.} 
This \textit{entanglement coupled cluster theory} is \revHK{outlined}, and proof-of-concept calculations for doublet states are reported.
The approach is \revHK{further} extendable to open-shell systems with other values of the total spin. 

\end{abstract}

\maketitle

\section{Introduction}

\revS{The interpretation of numerous spectroscopic experiments requires an exact treatment of electron spin in order to achieve satisfactory accuracy.
As the non-relativistic molecular electronic Hamiltonian commutes with the square of the total spin ($\vec{S}^2$) and the projected spin (${S}_z$), a common set of eigenfunctions exist for these operators.
Electronic structure methods that enforce the proper spin symmetries are called spin-adapted models.
Unfortunately, the formulation of accurate spin-adapted models for open-shell systems \sarai{is} complicated.
Consequently, the requirement that the wave function is an eigenfunction of $\vec{S}^2$ is often relaxed. Examples are the unrestricted Hartree-Fock (UHF) and spin-unrestricted (spin-orbital) coupled cluster theory.}

The complication with spin-unrestricted methods is the appearance of spin contamination.  
For UHF, spin contamination appears in two different situations.\citep{krylov2000} For open-shell systems, it arises from differences in the mean field experienced by the $\alpha$-- and $\beta$--electrons. For closed-shell systems, it appears when static correlation becomes important, for instance, when bonds are stretched \revS{or broken}.\citep{andrews1991,krylov2000,helgaker2014}
While spin contamination can be significant at low levels of theory, it is \revS{significantly} reduced as the description of the electronic structure improves. \citep{stanton1994,krylov2000} 
In the limit of full configuration interaction (FCI), there is no spin contamination.

\revS{Some molecular properties} are highly sensitive to the appearance of spin contamination and where
a wave function with the correct spin properties is desirable. For instance, this was demonstrated for the X-ray spectroscopy of the benzene cation\citep{vidal2020interplay}, and in general
in the description of magnetic properties.\citep{jost2013spin} Aside from the use of spin-adapted approaches, 
the problem of spin contamination has motivated the development of methods where contamination is reduced or where the spin properties are imposed in an average way.\citep{rittby1988open, knowles1993coupled,knowles2000erratum,neogrady1994spin,szalay1997spin, tsuchimochi2011constrained,biktagirov2020spin}

Spin adaptation is trivial for closed-shell species. In Hartree-Fock theory, a single determinant with doubly occupied molecular orbitals is an eigenfunction of both \revS{$\vec{S}^2$} and \revS{${S}_z$}. Post-Hartree-Fock methods can be parameterized in terms of singlet excitations of the closed-shell Hartree-Fock reference, thereby producing pure singlet approximate wave functions.\citep{helgaker2014} All singlet excitation operators with respect to the closed-shell reference determinant commute. In coupled cluster theory, this ensures that the Baker-Campbell-Hausdorff (BCH) expansion of the similarity-transformed Hamiltonian truncates after \revS{(}at most\revS{)} four nested commutators with the cluster operator. 

For open-shell systems, spin adaptation becomes more complicated. At {the} Hartree-Fock level, the restricted open-shell (ROHF) variant provides \sarai{a spin-adapted wave function}. 
However, including dynamical correlation with coupled cluster theory in a spin-adapted manner has \revS{turned out} to be a significant challenge.
As for closed-shell states,
a spin-adapted formulation of coupled cluster theory relies on \revS{expressing} the cluster operator in terms of singlet excitation operators (also called unitary group generators).
For open-shell systems, this implies the inclusion of singlet excitation operators where electrons are both created and annihilated in the singly occupied (active) orbitals. Such excitation operators do not commute and the BCH expansion for the similarity-transformed Hamiltonian no longer truncates after four nested commutators. 
Furthermore, the formulation of the cluster operator is not unique.\cite{li1994automation,herrmann2020generation}

One of the first implementations of spin-adapted coupled cluster theory for high-spin open-shell systems was presented by Janssen and Schaefer.\citep{janssen1991automated} Their choice of the cluster operator lacks terms necessary to span the full spin space for some spatial configurations.\citep{li1994automation,herrmann2020generation}
A similar approach is the unitary group approach developed by Li, Paldus, Jeziorski, and Jankowski.\citep{li1994automation,li1995spin,li1995unitary,jeziorski1995unitary,jankowski1999unitary} \revS{They criticized the cluster operator used by Janssen and Schaefer, which generates a linearly dependent basis for the coupled cluster equations.} This leads to an ambiguity in the formulation of the theory and convergence problems.\cite{li1994automation} In the unitary group approach, the cluster operator is defined such that the corresponding basis for the cluster amplitude equations is linearly independent and orthonormal.

Recently, Herrmann and Hanrath\citep{herrmann2020generation,herrmann2022correctly} present the automatic generation of a set of excitation operators for spin-adapted open-shell coupled cluster theory. The resulting basis for the coupled cluster equations is linearly independent but non-orthogonal. They arrive at a cluster operator (see Ref.~\citenum{herrmann2022correctly} for CCSD) that can be used for arbitrary high-spin open-shell systems. However, the operator (and its construction) is significantly more complicated than in the closed-shell theory, as is the case in the unitary group approach. 

The spin-adapted open-shell coupled cluster methods described so far are complicated by non-commuting contributions to the cluster operator, and by the complexity of the operator itself. 
\revS{Several authors\citep{nooijen1996many,nooijen1996general,nooijen2001towards,datta2008compact,datta2013non,datta2015communication} have advocated the use of normal ordering of the exponential operator to circumvent the problems arising from non-commuting terms. 
Nevertheless, the complexity of spin-adapted open-shell coupled cluster theory makes manual derivation and implementation impractical and automated generation of equations and code is seemingly inevitable.\citep{janssen1991automated,li1994automation,nooijen2001towards,datta2013non,herrmann2020generation, herrmann2022correctly}}


In this paper, we explore a novel strategy to obtain a spin-adapted description \revS{for open-shell systems}. 
In entanglement coupled cluster theory,
orbitals of the molecular system are mixed with orbitals from a non-interacting electron bath. A closed-shell determinant is constructed in this mixed orbital basis and used as a reference for spin-adapted closed-shell coupled cluster theory.
The cluster operator is defined as in standard spin-adapted closed-shell theory. That is, in terms of singlet excitation operators which all commute with each other. \revS{Hence, the BCH expansion of the similarity transformed Hamiltonian truncates after four nested commutators.}
The desired state of the molecular system is obtained by applying a projection operator that enforces the spin properties of the non-interacting bath. Due to the coupling of the system and the bath---into a \revS{singlet} state---this projection \revS{also} imposes restrictions on the molecular system. 

Compared to other approaches, the entanglement coupled cluster approach is simple in its formulation. 
\revS{Still}, the projection operator introduces significant complexity \revS{in} the working equations. The projection operator commutes with the Hamiltonian and the equations can be recast as a change of the projection manifold for the standard \revS{closed-shell} coupled cluster equations. The entanglement coupled cluster equations have a non-unit metric, \revS{i.e.,} the matrix elements of the projection operator in the basis defined by the cluster operator. This metric is rank-deficient, and linear dependencies in the basis must be removed. \revS{However, this is straightforward.}

In the following, we present the entanglement coupled cluster theory and its application to doublet states. We describe the ground state equations and the extension to equation-of-motion for excited states. We also outline how triplet systems can be described within the same framework. Finally, we present proof-of-concept calculations for the CCS and CCSD variants of the theory. 

\section{General formulation of entanglement coupled cluster theory}
We consider the prospects of exploiting the entanglement of two subsystems---\revS{the} molecule and a fictitious non-interacting bath---to describe open-shell systems with \revS{spin-adapted} coupled cluster \revS{theory}.
\revS{W}e formulate the theory for doublet systems, but we will also outline the extension to high-spin triplets. 

\subsection{The molecular system, the bath, and a mixed orbital basis}
\revS{The} molecule ($\mathrm{m}$) and a non-interacting electron bath ($\mathrm{b}$) are coupled to a singlet spin state ($S=0$). \revS{The Hamiltonian of the total system is given by
\begin{align}
    H = H_\mathrm{m} + H_\mathrm{b},
\end{align}
and the following commutator relations hold for the projected spin, squared spin, and number operators:}
\begin{align}
   & [H, N_{\mathrm{m}}] = [H, N_{\mathrm{b}}] = [H, N] = 0\label{eq:commutator1}\\
   & [H, \vec{S}_{\mathrm{m}}^2] = [H, \vec{S}_{\mathrm{b}}^2] = [H, \vec{S}^2]  = 0\label{eq:commutator2}\\
   & [H, S_{\mathrm{m}z}] = [H, S_{\mathrm{b}z}] = [H, S_{z}]  = 0.\label{eq:commutator3}
\end{align}

From the addition theorem of angular momentum, we know that two angular momenta can only couple to zero total angular momentum if they have equal magnitude\sarai{. That} is, $S_\mathrm{m} = S_\mathrm{b}$, such that $S = |S_\mathrm{m} - S_\mathrm{b}| = 0$.

\revS{Eigenfunctions of $H$ can be chosen as eigenfunctions of ${\vec{S}}^2$ and $S_z$.} 
Using the Clebsch-Gordan coefficients, we may expand the eigenfunctions of ${\vec{S}}^2$ and $S_z$  \revS{in} the product basis of eigenfunctions of $\vec{S}_{\mathrm{m}}^2$ and $S_{\mathrm{m}z}$, and $\vec{S}_{\mathrm{b}}^2$ and $S_{\mathrm{b}z}$. If we only consider systems with $S=0$ \sarai{(}$\ket{S\; M} = \ket{0\; 0}$\sarai{)}, we obtain
\begin{align}
\begin{split}
    \ket{0\; 0}
    &= \sum_{\gamma\delta}\sum_{s m} \ket{\gamma\; s m}\otimes \ket{\delta\; s-m}C^{s,s,0}_{m,-m,0}\Gamma_{\gamma\delta},\label{eq:CG}
\end{split}
\end{align}
where we let the indices $\gamma$ and $\delta$ capture all characteristics of the states in the uncoupled picture (apart from their spin)\revS{. The coefficient $\Gamma_{\gamma\delta}$ represents the correlation between the states in the uncoupled picture. T}he $C^{s,s,0}_{m,-m,0}$ are the Clebsch-Gordan coefficients.

\revS{In the following, we consider a total system with $N_e$ electrons and a bath with a single spatial orbital ($\phi_{\mathcal{B}}$). Since the total system is a singlet, $N_e = N_\mathrm{m} + N_\mathrm{b}$ is even.
The theory can be extended to include more orbitals in the bath.}

The standard non-relativistic electronic Hamiltonian is used\revS{:}
\begin{align}
\begin{split}
    H &= \sum_{pq} {h}_{pq} E_{pq} + \frac{1}{2}\sum_{pqrs}{g}_{pqrs}(E_{pq}E_{rs} - \delta_{qr}E_{ps})\\
    &+ {h}_{\mathcal{B}}E_{\mathcal{BB}} + \frac{1}{2}{g}_{\mathcal{B}}(E_{\mathcal{BB}}E_{\mathcal{BB}} - E_{\mathcal{BB}}), \label{eq:hamiltonian}
\end{split}
\end{align}
where 
\begin{align}
    E_{pq} = c^\dagger_{p\alpha}c_{q\alpha} + c^\dagger_{p\beta}c_{q\beta} = E_{pq}^{\alpha} + E_{pq}^{\beta}
\end{align}
is a singlet excitation operator, and $c^\dagger_{p\sigma}$ and $c_{p\sigma}$ respectively create and annihilate a $\sigma$-spin electron in spatial orbital $\phi_p$.
Since the molecule does not interact with the bath, the summations in Eq. \eqref{eq:hamiltonian} are restricted to the molecular orbitals\sarai{. The} ${h}_{\mathcal{B}}$ and ${g}_{\mathcal{B}}$ determine the one- and two-electron interactions within the bath. 
\begin{figure*}
    \centering
    \includegraphics[width=0.90\linewidth,trim={0 1.8cm 0 1cm},clip]{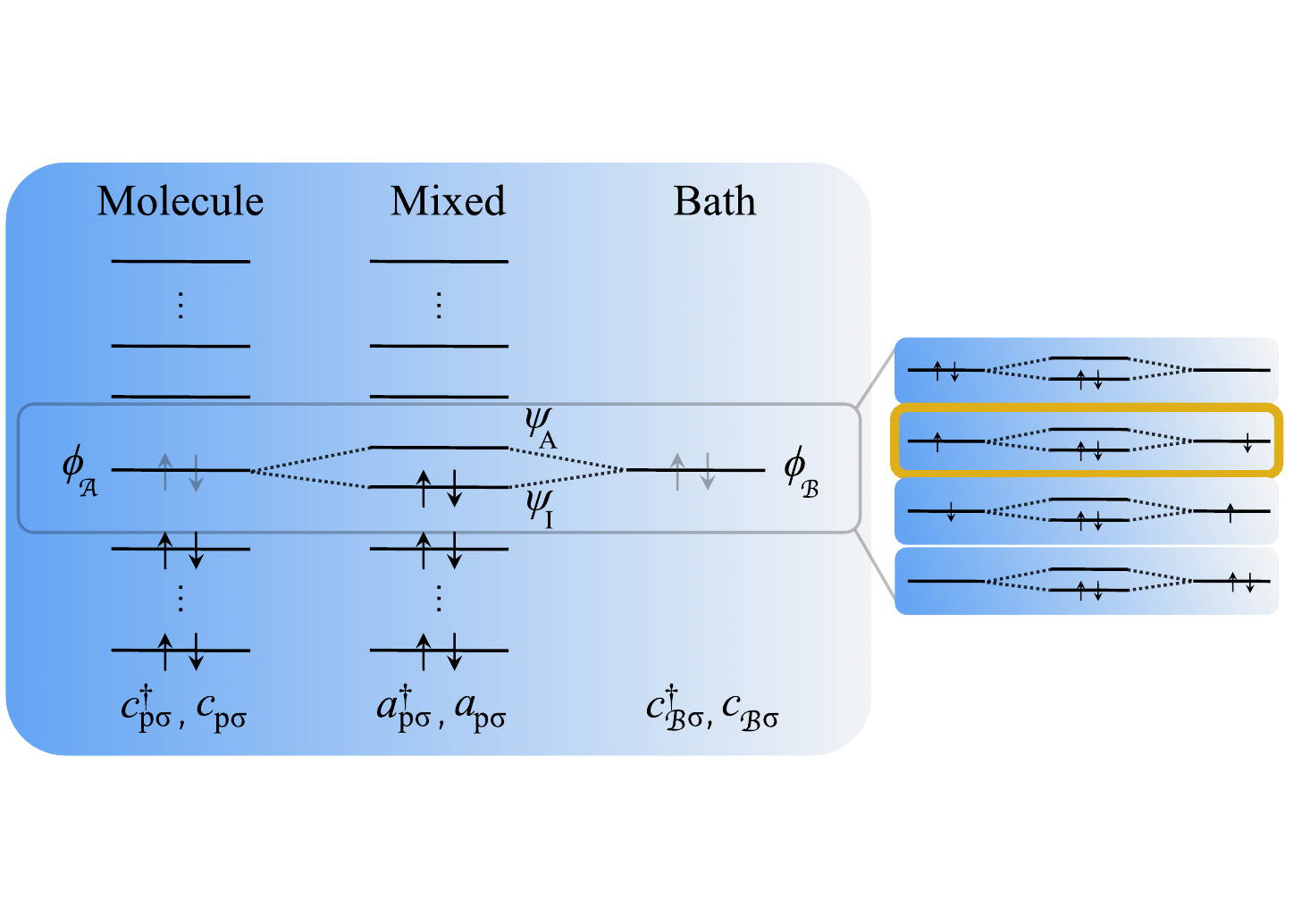}
    \caption{Illustration of the molecular reference, the non-interacting bath, and the resulting mixed orbital reference determinant. Creation and annihilation operators $c^\dagger_{p\sigma}$, $c^\dagger_{\mathcal{B}\sigma}$, and $c_{p\sigma}$ and $c_{\mathcal{B}\sigma}$ refer to the original basis, whereas $a^\dagger_{p\sigma}$ and $a_{p\sigma}$ refer to the mixed orbital basis. \revS{The mixed closed-shell reference is a linear combination of configurations with 2, 1, or 0 electrons in the active orbital of the molecule, see Eqs. \eqref{eq:one-orbital-system-2} and \eqref{eq:one-orbital-system-3}. One of these configurations has a single $\alpha$-spin electron in the active molecular orbital (highlighted in yellow).}}
    \label{fig:msb}
\end{figure*}


The molecular orbitals are divided into three disjoint sets: the doubly occupied core orbitals, $\{\phi_i^{\mathcal{C}}\}$, a single active orbital, $\{\phi_{\mathcal{A}}\}$, and the virtual orbitals, $\{\phi_a^{\mathcal{V}}\}$.
\revS{We introduce an orthogonal} transformation that mixes the bath orbital and the active orbital,
\begin{align}
\psi_I &= \phi_{\mathcal{A}}\cos\theta - \phi_{\mathcal{B}}\sin\theta\label{eq:occ_mixed}\\
\psi_A &=  \phi_{\mathcal{A}}\sin\theta +  \phi_{\mathcal{B}}\cos\theta.\label{eq:vir_mixed}
\end{align}
\revS{This transformation defines the mixed orbital basis, given in terms of a mixing angle $\theta$.} In Fig. \ref{fig:msb}, we illustrate the formation of the mixed orbital basis\revS{.}
\revS{Unless otherwise stated,} we will use indices $i, j, k$ to denote core orbitals, $a, b, c$ to denote virtual orbitals, $I$ and $A$ to denote the orbitals resulting from mixing the active and bath orbitals, and $p, q, r, s$ to denote general orbitals. 

We can define a transformation matrix $\boldsymbol{{U}}$ acting on the full set of orbitals (core, active, bath, and virtual),
\begin{align}
\boldsymbol{{U}}
    =\begin{pmatrix}
    \boldsymbol{I} & \boldsymbol{0} & \boldsymbol{0}\\
    \boldsymbol{0}  & \boldsymbol{u} &\boldsymbol{0}\\
     \boldsymbol{0} & \boldsymbol{0}  & \boldsymbol{I}\\
    \end{pmatrix},
\end{align}
where 
\begin{align}
   \boldsymbol{{u}}
    =\begin{pmatrix}
    \cos\theta & -\sin\theta \\
    \sin\theta & \cos\theta
    \end{pmatrix}, \label{eq:theta_transform}
\end{align}
such that 
\begin{align}
\boldsymbol{\psi} =
    \boldsymbol{{U}}
        \begin{pmatrix}
    \boldsymbol\phi^{\mathcal{C}}\\
    \phi_{\mathcal{A}}\\
    \phi_{\mathcal{B}}\\
    \boldsymbol\phi^{\mathcal{V}}\\
    \end{pmatrix}.
\end{align}
In the resulting mixed orbital basis, the core and virtual orbitals remain unchanged, whereas the active and bath orbitals are mixed. The transformed orbitals are orthonormal since the transformation $\boldsymbol{U}$ is \revS{orthogonal}.

\subsection{The reference state}
The transformation of the orbitals corresponds to a transformation of the creation and annihilation operators,
\begin{align}
    {a}^{\dagger}_{p\sigma} = \sum_{r}{U}_{pr}{c}^{\dagger}_{r\sigma}.
\end{align}
\revS{We now consider a closed-shell reference determinant where $\{\phi^{\mathcal{C}}_i\}$ and $\psi_I$ are doubly occupied and where $\{\phi^{\mathcal{V}}_a\}$ and $\psi_A$ are unoccupied.} We have
\begin{align}
\begin{split}
        \ketR &= {a}^\dagger_{I\alpha}{a}^\dagger_{I\beta}\prod_{i}{a}^\dagger_{i\alpha}{a}^\dagger_{i\beta}
    \ket{\text{vac}}\\
    &= {a}^\dagger_{I\alpha}{a}^\dagger_{I\beta}
    \prod_{i}{c}^\dagger_{i\alpha}{c}^\dagger_{i\beta}\ket{\text{vac}},\label{eq:reference_state}
\end{split}
\end{align}
where $\ket{\text{vac}}$ denotes the true vacuum. 
\revS{For simplicity,} we consider \revS{the case} where the molecule only has a single orbital, which is active. In the transformed basis\sarai{,} there are two orbitals ($\psi_I, \psi_{A}$) and the reference determinant becomes
\begin{equation}
    \ketR = 
    {a}^\dagger_{I\alpha}{a}^\dagger_{I\beta}\ket{\text{vac}}.\label{eq:one-orbital-system-1}
\end{equation}
We \revS{now} express this state in the original basis\revS{,}
\begin{equation}
\begin{split}
    \ketR = 
    \Big(&\cos^2\theta\;{c}^\dagger_{\mathcal{A}\alpha}{c}^\dagger_{\mathcal{A}\beta}
    + \sin^2\theta\;{c}^\dagger_{\mathcal{B} \alpha}{c}^\dagger_{\mathcal{B}\beta}\\
    - &\cos\theta\sin\theta(
    {c}^\dagger_{\mathcal{A} \alpha}{c}^\dagger_{\mathcal{B}\beta} 
    -{c}^\dagger_{\mathcal{A}\beta}{c}^\dagger_{\mathcal{B}\alpha})\Big)\ket{\text{vac}},\label{eq:one-orbital-system-2}
\end{split}
\end{equation}
\revS{and obtain a linear combination of the three singlet states obtained from two electrons in two orbitals.}
Comparing to Eq. \eqref{eq:CG}, and using the notation $\ket{N _\mathrm{x}sm}_\mathrm{x},\;\mathrm{x}\in\{\mathrm{m}, \mathrm{b}\}$, we can rewrite Eq. \eqref{eq:one-orbital-system-2} as
\begin{align}
\begin{split}
    \ketR =&\Big(\ket{200}_\mathrm{m}\otimes\ket{000}_\mathrm{b}\Big)\cos^2\theta  \\ 
    + &\Big(\ket{000}_\mathrm{m}\otimes\ket{200}_\mathrm{b}\Big) \sin^2\theta \\
    -& \Big(\Ket{1\tfrac{1}{2}\tfrac{1}{2}}_\mathrm{m}\otimes\Ket{1  \tfrac{1}{2} \text{-}\tfrac{1}{2}}_\mathrm{b}\Big)\cos\theta\sin\theta\\
    -&\Big( \Ket{1  \tfrac{1}{2} \text{-}\tfrac{1}{2}}_\mathrm{m}\otimes\Ket{1  \tfrac{1}{2} \tfrac{1}{2}}_\mathrm{b}\Big)\cos\theta\sin\theta,
\end{split}
\label{eq:one-orbital-system-3}
\end{align}
with $\gamma = N_\mathrm{m}$ and $\delta = N_\mathrm{b}$.
\revS{T}he molecular system and the bath are entangled, i.e., the state of (or absence of) electrons in the bath orbital uniquely determines the state of electrons in the molecular system. Together they satisfy the condition for two electrons coupled to a singlet spin state. 

\subsection{Projection operators}
We introduce \revS{the} operator\revS{s}
\begin{align}
    P^\sigma = {c}^\dagger_{\mathcal{B}\sigma}{c}_{\mathcal{B}\sigma} = E_{\mathcal{BB}}^{\sigma},\label{eq:Ps_def}
\end{align}
that count the number of $\sigma$ electrons in the bath orbital (0 or 1). If, for example, we let $ P^\beta$ act on the state given by Eq. \eqref{eq:one-orbital-system-2}, we \revS{eliminate} all terms in the linear combination which do not have a $\beta$-electron in the bath orbital:
\begin{equation*}
\begin{split}
    P^\beta\ketR=\big(&\sin^2\theta\;{c}^\dagger_{\mathcal{B} \alpha}{c}^\dagger_{\mathcal{B} \beta}
    - \cos\theta\sin\theta
    {c}^\dagger_{\mathcal{A} \alpha}{c}^\dagger_{\mathcal{B} \beta} 
   \big)\ket{\text{vac}}.
\end{split}
\end{equation*}
We also introduce the operator\revS{s}
\begin{align}
    Q^\sigma = 1 - P^\sigma,\label{eq:Qs_def}
\end{align}
\sarai{that} count the number of $\sigma$-electron holes in the bath orbital (0 or 1).
Returning to the \sarai{minimal} example, we see that different combinations of $P^\sigma$ and $Q^\tau$ will extract different terms from the linear combination in Eq. \eqref{eq:one-orbital-system-2}:
\begin{align}
\begin{split}
    P^{\alpha}P^{\beta}\ketR
    &={c}^\dagger_{\mathcal{B} \alpha}{c}^\dagger_{\mathcal{B} \beta}
   \ket{\text{vac}}\sin^2\theta
\end{split}\label{eq:PP}\\
\begin{split}
    Q^{\alpha}Q^{\beta}\ketR
   &= {c}^\dagger_{\mathcal{A} \alpha}{c}^\dagger_{\mathcal{A} \beta}
   \ket{\text{vac}}\cos^2\theta
\end{split}\label{eq:QQ}\\
\begin{split}
    \PQb\ketR
   &=  
    {c}^\dagger_{\mathcal{A} \beta}{c}^\dagger_{\mathcal{B} \alpha}\ket{\text{vac}}\cos\theta\sin\theta
\end{split}\label{eq:PQ_1}\\
\begin{split}
    P^{\beta}Q^{\alpha}\ketR
    &=
    {c}^\dagger_{\mathcal{A} \alpha}{c}^\dagger_{\mathcal{B} \beta}\ket{\text{vac}}( - \cos\theta\sin\theta).\label{eq:doublet-spin-projector}
\end{split}
\end{align}
These product operators are orthogonal projection operators and satisfy 
\begin{align}
\begin{split}
    \P^\dagger &= \P\\
    \P ^2 &= \P\\
    \P \P ' &= 0,\;\;\P  \neq \P ' ,
\end{split}\label{eq:projection_prop}
\end{align}
where $\P , \P ' \in \{ P^{\alpha}P^{\beta}, Q^{\alpha}Q^{\beta}, \PQb, P^{\beta}Q^{\alpha}\}$.
Note that the operators \revS{only} refer to the bath orbital. It is through the particular coupling between the bath and system that we obtain specific states of the system upon application of $\P$.
We will use the closed-shell determinant defined in Eq. \eqref{eq:reference_state} as a reference for coupled cluster theory. We will then apply projection operators to enforce properties on the system. \revS{This is the entanglement coupled cluster (ECC) approach.}
\revS{In the following section, we briefly review the standard spin-adapted closed-shell coupled cluster theory to establish the notation used to describe ECC.}

\subsection{Spin-adapted closed-shell coupled cluster theory}
In \sarai{coupled cluster theory, the wave function} is given by\citep{helgaker2014}
\begin{align}
   \ket{\mathrm{CC}} = \exp(T)\ketR, 
\end{align}
where $\ketR$ is a \sarai{reference} determinant (typically the restricted Hartree-Fock determinant), and
\begin{align}
    T = T_1 + T_2 + \cdots
\end{align}
is the cluster operator. \revS{$T_1$ and $T_2$ generate single and double excitations} of the reference, and so on. 
In the spin-adapted closed-shell theory, \sarai{the reference determinant is closed-shell and} the cluster operator is defined in terms of singlet excitation operators\sarai{:} 
\begin{align}
    A_{pq} = a^\dagger_{p\alpha}a_{q\alpha} + a^\dagger_{p\beta}a_{q\beta} = A_{pq}^{\alpha} + A_{pq}^{\beta}.
\end{align}
\sarai{F}or $T_1$ and $T_2$ we have
\begin{align}
    T_1 = \sum_{\mu_1}t_{\mu_1}\tau_{\mu_1} = \sum_{ai} t_{i}^{a}A_{ai}\label{eq:t1}
\end{align}
and
\begin{align}
    T_2 = \sum_{\mu_2}t_{\mu_2}\tau_{\mu_2} = \frac{1}{2}\sum_{aibj} t_{ij}^{ab}A_{ai}A_{bj}.\label{eq:t2}
\end{align}
The $\boldsymbol{t}$ parameters are the cluster amplitudes. The truncation of {$T$} yields the different standard coupled cluster models: CCS with $T=T_1$, CCSD with $T=T_1 + T_2$, etc.

The ground state coupled cluster equations are obtained by projecting the Schrödinger equation onto a set of vectors $\{\braR\exp(-T), \bra{\mu}\exp(-T)\}$, where $\bra{\mu} = \braR \tau_{\mu}^\dagger$:
\begin{align}
\begin{split}
    E_0 &= \Tbraket{\mathrm{R}}{\simH}{\mathrm{R}} \\
    \Omega_{\mu} &= \Tbraket{\mu}{\simH}{\mathrm{R}} = 0.\label{eq:omega_standard_cc}
\end{split}
\end{align}
Here, we have introduced the similarity-transformed Hamiltonian ${\simH} = \exp(-T)H\exp(T)$. The first of these equations gives the energy\revS{, and} the second set of equations must be solved to determine the cluster amplitudes. 

Excited states are obtained through linear response theory or the equation-of-motion (EOM) approach.
In EOM coupled cluster theory, the states $\ket{k}$ are defined by the expansion
\begin{align}
    \ket{k}=\sum_{\mu\geq0}\exp({T})R^k_{\mu}\ket{\mu},
\end{align}
{where $\ket{\mu} = \tau_{\mu}\ketR$},
and $\boldsymbol{R^k}$ are the right eigenvectors
of the similarity-transformed Hamiltonian:
\begin{align}
   \boldsymbol{\simH} \boldsymbol{R}^k = E_k \boldsymbol{R}^k.
\end{align}
The similarity-transformed Hamiltonian has the form
\begin{align}
    \boldsymbol{\simH} =
    \begin{pmatrix}
     E_0 & \boldsymbol{\eta}^\mathrm{T} \\
     \boldsymbol\Omega &\boldsymbol{J} + E_0\boldsymbol{I} \\
    \end{pmatrix}=
        \begin{pmatrix}
     E_0 & \boldsymbol{\eta}^\mathrm{T} \\
     \boldsymbol 0 & \boldsymbol{J} + E_0\boldsymbol{I} \\
    \end{pmatrix},\label{eq:shape_of_simH}
\end{align} 
where $\boldsymbol{J}$ is the Jacobian matrix with elements 
\begin{align}
    J_{\mu\nu}=\Tbraket{\mu}{[\simH,\tau_{\nu}]}{\mathrm{R}},
\end{align}
and
\begin{align}
    \eta_{\nu} = \Tbraket{\mathrm{R}}{[\simH, \tau_{\nu}]}{\mathrm{R}}.
\end{align} 
In Eq. \eqref{eq:shape_of_simH}, we have assumed that the ground state equations in \eqref{eq:omega_standard_cc} are solved, such that $\boldsymbol{\Omega} = \boldsymbol{0}$.
The eigenvalues of $\boldsymbol{\simH}$ are the energies of the electronic states in EOM coupled cluster theory, and the excitation energies $\omega_k$ are the eigenvalues of $\boldsymbol{J}$.

Since $\boldsymbol{\simH}$ is non-Hermitian, its left and right eigenvectors differ. We may express the left EOM coupled cluster states as
\begin{align}
   \bra{k} = \sum_{\mu\geq0}L^k_{\mu}\bra{\mu}\exp(-T),
\end{align}
where
\begin{align}
   \boldsymbol{\simH}^T {\boldsymbol{L}^{k}} = E_k {\boldsymbol{L}^{k}},
\end{align}
and we require that the left and right states form a biorthonormal set:
\begin{align}
    \Dbraket{k}{l} = \delta_{kl}.\label{eq:biorth}
\end{align}

The right vectors are given by
\begin{align}
    \boldsymbol R^0 = 
    \begin{pmatrix}
    1\\
    0\\
    \end{pmatrix}, \; \;  
    \boldsymbol R^k = 
    \begin{pmatrix}
    \omega_k^{-1}\boldsymbol{\eta}^\mathrm{T}\boldsymbol{r}_k \\
    \boldsymbol{r}_k\\
    \end{pmatrix}\; \text{for}\; k>0, \label{eq:right_std}
\end{align}
where $\boldsymbol{r}_k$ are the right eigenvectors of $\boldsymbol{J}$, corresponding to the eigenvalue $\omega_k$. The first element of $\boldsymbol{R}_k,\; k>0$ is obtained from the biorthonormalization condition in Eq. \eqref{eq:biorth}.
The left vectors are given by
\begin{align}
    \boldsymbol L^0 = 
    \begin{pmatrix}
    1 \\
    \boldsymbol{\bar{t}}\\
    \end{pmatrix}, \; \; 
    \boldsymbol L^k = 
    \begin{pmatrix}
    0 \\
    \boldsymbol{l}_k\\
    \end{pmatrix}\;\text{for}\; k>0,\label{eq:left_std}
\end{align}
where $\boldsymbol{\bar{t}}$ are the left ground state amplitudes, determined by solving 
\begin{align}
    \boldsymbol{J}^\mathrm{T}\boldsymbol{\bar{t}} = -\boldsymbol{\eta},\label{eq:multipliers}
\end{align}
and $\boldsymbol{l}_k$ is a left eigenvector of $\boldsymbol{J}$, corresponding to the eigenvalue $\omega_k$.

\subsection{Entanglement coupled cluster theory\label{sec:ECC}}
We will use the closed-shell determinant given in Eq. \eqref{eq:reference_state} as our reference to define a coupled cluster wave function. The cluster operator is defined in the transformed basis \revS{(}in terms of $a^\dagger_{p\sigma}$ and $a_{p\sigma}$\revS{)} and in the same way as in spin-adapted closed-shell theory\revS{; see the definitions of $T_1$ and $T_2$ in Eqs. \eqref{eq:t1} and \eqref{eq:t2}, but note that the summations will include the mixed \revS{orbital indices $I$ and $A$.} The resulting coupled cluster state is a pure singlet state, and, similar to the mixed orbital reference, it is a linear combination of configurations with $N_m = N_e$, $N_m = N_e - 1$, and $N_m = N_e - 2$.}

To describe a particular state of the molecular system, we apply a projection operator $\P $ that enforces conditions on the bath. The desired state of the system is imposed through its coupling to the bath. The projected coupled cluster wave function is
\begin{align}
   \ket{\mathrm{ECC}} = \P \exp(T)\ketR.
\end{align}

The Hamiltonian in Eq. \eqref{eq:hamiltonian} and the projection operators $\P$ in Eqs. \eqref{eq:PP}--\eqref{eq:doublet-spin-projector} are defined in terms of the original creation and annihilation operators $c^\dagger_{p\sigma}$ and $c_{p\sigma}$. 
Before solving the coupled cluster equations, we must transform $H$ and $\P $ to the mixed orbital basis. For the Hamiltonian, this amounts to a transformation of the one- and two-electron integrals,
\begin{align}
    H=\sum_{pq} \tilde{h}_{pq} A_{pq} + \frac{1}{2}\sum_{pqrs}\tilde{g}_{pqrs}(A_{pq}A_{rs} - \delta_{qr}A_{ps}),\label{eq:hamiltonian-2}
\end{align}
where
\begin{align}
    \tilde{h}_{pq} &= \sum_{tu} U_{pt}{h}_{tu}U_{qu}\\
     \tilde{g}_{pqrs} &= \sum_{tuvw} U_{pt} U_{qu}{g}_{tuvw} U_{rv} U_{sw}.
\end{align}
For the projection operators $P^{\sigma}$ in Eq. \eqref{eq:Ps_def}, we obtain
\begin{align}
\begin{split}
    P^{\sigma} =&\phantom{-}\sin^2\theta A_{II}^{\sigma} + \cos^2\theta A_{AA}^{\sigma}\\
    &- \cos\theta\sin\theta(A^{\sigma}_{IA} + A^{\sigma}_{AI}).
\end{split}
\end{align}

By pre-multiplying the Schrödinger equation with $\exp(-T)\mathcal{P}$ and projecting onto the vectors $\{\braR, \bra{\mu}\}$, we obtain the equations for the ground state energy and amplitudes\revS{:}
\begin{align}
    \Tbraket{\mathrm{R}}{\simP\simH}{\mathrm{R}} &= E_0 \Tbraket{\mathrm{R}}{\simP}{\mathrm{R}}\label{eq:ECC-omega}\\
    \Omega_\mu &= {\Omega}^S_\mu E_0, \label{eq:ECC-E}
\end{align}
where 
\begin{align}
    \Omega_\mu  &= \Tbraket{\mathrm{\mu}}{\simP\simH}{\mathrm{R}}\\
    \Omega^S_\mu &= \Tbraket{\mu}{\simP}{\mathrm{R}}.
\end{align}
Here, we have \sarai{used} the resolution of the identity $\exp(-T)\exp(T) = 1$, \revS{introduced} $\simP = \exp(-T)\P \exp(T)$, and \revS{used} $[H,\P] = 0$ and $\P^2=\P$.
Compared to standard coupled cluster theory, a non-unit metric enters the right-hand sides of Eqs. \eqref{eq:ECC-omega} and \eqref{eq:ECC-E}. 
The equations can be viewed as a change of the projection manifold in the standard theory to
$\{\braR \simP, \bra{\mu}\simP\}$.

To derive the working equations, we evaluate $\simH\ketR$ and the projection onto $\{\braR \simP, \bra{\mu}\simP\}$. Due to the definition of $T$, the BCH expansion of $\simH$ truncates after four nested commutators in general. With a $T_1$-transformed $H$, maximally three nested commutators with $T_2$ enter the ECCSD equations, since $\bra{\mu_2}\simP$ contains quadruply excited determinant\revS{s} $\bra{\mu_2}E_{IA}^{\beta}E_{IA}^{\alpha}$.

The basis $\{\bra{\mu}\simP\}$ can have redundancies that must be removed to solve the ECC equations uniquely. We have found these redundancies by diagonalizing the matrix
\begin{align}
  {J}^S_{\mu\nu} &= \Tbraket{\mu}{\simP}{\nu}  
\end{align}
and analyzing the null space. There are two cases:
\begin{enumerate}
    \item $\bra{\mu}\simP = 0$, in which case $\bra{\mu}$ is removed from the projection space and $\tau_\mu$ is removed from $T$.
    \item \revS{The v}ectors $\{\bra{\mu}\simP\}$ are linearly dependent, in which case we remove the appropriate number of vectors and corresponding excitation operators, preferring to remove those of higher excitation order.
\end{enumerate}

The overall scaling of solving the ECCSD ground state equations, Eq. \eqref{eq:ECC-omega}, is $\mathcal{O}(N^6)$, because any contributions to the vectors $\{\braR \simP, \bra{\mu}\simP\}$ from excited determinants of \revS{excitation} order three and four have restricted indices (see Appendix \ref{sec:equations}). These higher order determinants result in additional costs compared to CCSD, but not higher scaling. All contributions from singly and doubly excited determinants can be implemented at the same cost as standard CCSD (with some extra $\mathcal{O}(N^4)$ operations).

The ECC excited states are obtained within the EOM framework. We obtain the generalized eigenvalue equations
\begin{align}
    \boldsymbol{\simH R}^k = E_k\boldsymbol{\simS R}^k\label{eq:right}\\
    \boldsymbol{\simH}^T\boldsymbol{L}^k = E_k\boldsymbol{\simS}^T\boldsymbol{L}^k,\label{eq:left}
\end{align}
where 
\begin{align}
\boldsymbol{\simH}=
    \begin{pmatrix}
        \Tbraket{\mathrm{R}}{\simP\simH}{\mathrm{R}} &\boldsymbol{{\eta}}^\mathrm{T}\\
        \boldsymbol{{\Omega}} &  \boldsymbol{J}
    \end{pmatrix},
\end{align}
and
\begin{align}
    \boldsymbol{\simS} =
        \begin{pmatrix}
        \Tbraket{\mathrm{R}}{\simP}{\mathrm{R}} &{\boldsymbol{\eta^S}}^\mathrm{T}\\
       \boldsymbol{\Omega^S}& \boldsymbol{{J^S}}
    \end{pmatrix},
\end{align} 
and where we have introduced 
\begin{align}
\begin{split}
   \eta_\nu &=\Tbraket{\mathrm{R}}{\simP\simH}{\nu}\\
    J_{\mu\nu}&=\Tbraket{\mu}{\simP\simH}{\nu}\\
 {\eta}^S_{\nu} &= \Tbraket{\mathrm{R}}{\simP}{\mathrm{\nu}}.
\end{split}
\end{align}
The lowest generalized eigenvalue is the ground state energy and the remaining eigenvalues are excited state energies. The corresponding right and left eigenvectors have the same form as in standard closed-shell EOM theory, see Eqs. \eqref{eq:right_std} and \eqref{eq:left_std}. However, the $\boldsymbol{\bar{t}}$-equation is now given by 

\begin{align}
     (\boldsymbol{J}-E_0{\boldsymbol{J^S}})^\mathrm{T} \tbar =  E_0{\boldsymbol{\eta^S}} - \boldsymbol{\eta}.
\end{align}

\subsection{Size-extensivity of excited states}
In this section, we will analyze the scaling properties of the EOM entanglement coupled cluster energies. 
We consider two non-interacting systems, $A$ and $B$. \revS{System $A$ is open-shell and system $B$ is a \sarai{singlet}.}
Since the systems $A$ and $B$ do not interact, we have
\begin{align}
\begin{split}
    T &= T_A + T_B\\
    H &= H_A + H_B,\label{eq:extensivity_op}
\end{split}
\end{align}
and since only system $A$ is open-shell, we have
\begin{align}
    \P &= \P_A.\label{eq:extensivity_P}
\end{align}
We have the following commutator relations between the operators of systems $A$ and $B$:
\begin{align}
\begin{split}
   & [T_A, T_B] = 0\\
   & [H_A, T_B] = [H_B, T_A] = 0\\
   & [\P_A, T_B] = 0.\label{eq:extensivity_commutator}
\end{split}
\end{align}
The size-extensivity of the ground state follows directly from these commutator relations, as in the standard theory (see Ref.~\citenum{helgaker2014}).

\revS{Using the properties of the operators (Eqs. \eqref{eq:extensivity_op} and \eqref{eq:extensivity_P}) and the commutator relations (Eq. \eqref{eq:extensivity_commutator}), we obtain the block structure}
\begin{widetext}
\begin{align}
\begin{split}
\boldsymbol{\simH}  =
    \begin{pmatrix}
    \boldsymbol{\simH}_1 & \boldsymbol{\simH}_{2}\\
    \boldsymbol{0} & \boldsymbol{\simH}_{3}\\
    \end{pmatrix}
    =
    \begin{pmatrix}
    \simH_{0,0} & 
    \boldsymbol{\simH}_{0,A} & \boldsymbol{\simH}_{0,B} & \boldsymbol{\simH}_{0,AB}\\
    \boldsymbol{\simH}_{A,0} & \boldsymbol{\simH}_{A,A} & \boldsymbol{\simH}_{A,B} & \boldsymbol{\simH}_{A,AB}\\
    \boldsymbol{0} & 
    \boldsymbol{0} & \boldsymbol{\simH}_{B,B} & \boldsymbol{\simH}_{B,AB}\\
    \boldsymbol{0} 
    & \boldsymbol{0} & \boldsymbol{\simH}_{AB,B} & \boldsymbol{\simH}_{AB,AB}\\
    \end{pmatrix}
\end{split}
\end{align}
and
\begin{align}
\begin{split}
\boldsymbol{\simS} = 
    \begin{pmatrix}
    \boldsymbol{\simS}_1 & \boldsymbol{0}\\
    \boldsymbol{0} & \boldsymbol{\simS}_{3}\\
    \end{pmatrix}
    =
    \begin{pmatrix}
    \simS_{0,0} & 
    \boldsymbol{\simS}_{0,A} & \boldsymbol{0} & \boldsymbol{0}\\
    \boldsymbol{\simS}_{A,0} & \boldsymbol{\simS}_{A,A} & \boldsymbol{0} & \boldsymbol{0}\\
    \boldsymbol{0} & 
    \boldsymbol{0} & \boldsymbol{\simS}_{B,B} & \boldsymbol{\simS}_{B,AB}\\
    \boldsymbol{0} 
    & \boldsymbol{0} & \boldsymbol{\simS}_{AB,B} & \boldsymbol{\simS}_{AB,AB}\\
    \end{pmatrix},\\
\end{split}
\end{align}
\revS{in the basis
$\ket{\mu_A, \mu_B}$ = $\ket{\mu_A}\otimes\ket{\mu_B}$, for $\mu_X \geq 0$.}
\revS{The} subscript $0$ denotes the reference determinant for both systems $A$ and $B$. Subscripts $A$, $B$, or $AB$ denote excited determinants in system $A$, $B$, or both.
The characteristic equation for the generalized eigenvalue equation becomes
\begin{align}
\begin{split}
    &\det
    \begin{pmatrix}
    \boldsymbol{\simH}_1 - E\boldsymbol{\simS}_1 & \boldsymbol{\simH}_{2}\\
    \boldsymbol{0} & \boldsymbol{\simH}_{3}- E\boldsymbol{\simS}_3\\
    \end{pmatrix}
    = \det(\boldsymbol{\simH}_1 - E\boldsymbol{\simS}_1)\det(\boldsymbol{\simH}_{3}- E\boldsymbol{\simS}_3) = 0.
\end{split}
\end{align}
Thus, the generalized eigenvalues of $\{\boldsymbol{\simH},\boldsymbol{\simS}\}$ are the collected generalized eigenvalues of $\{\boldsymbol{\simH}_1, \boldsymbol{\simS}_1\}$ and $\{\boldsymbol{\simH}_3, \boldsymbol{\simS}_3\}$.
We start by considering $\boldsymbol{\simH}_1$ and $\boldsymbol{\simS}_1$. We have
\begin{align}
\begin{split}
  \boldsymbol{\simH}_1 &=
  \begin{pmatrix}
      \simH_{0,0} & 
    \boldsymbol{\simH}_{0,A}\\
    \boldsymbol{\simH}_{A,0} & \boldsymbol{\simH}_{A,A}
  \end{pmatrix} 
  =
  \begin{pmatrix}
      \simS^A_{0} E_{A} & 
     \boldsymbol{\simH}^A_{0,A}\\
    \boldsymbol{\simS}^{A}_{A, 0}E_A & \boldsymbol{\simH}^A_{A,A}
  \end{pmatrix} + E_B\boldsymbol{\simS}^{A}_1 ,
  \end{split}\label{eq:H1}
\end{align}
\begin{align}
  \boldsymbol{\simS}_1 = \begin{pmatrix}
      \simS^{A}_{0,0} & 
    \boldsymbol{\simS}^{A}_{0,A}\\
    \boldsymbol{\simS}^{A}_{A,0} & \boldsymbol{\simS}^{A}_{A,A}
  \end{pmatrix} = \boldsymbol{\simS}^{A}_1,
\end{align}
where we have used the relations
\begin{align}
   \simH_{0,0} =&\; \simS^{A}_{0,0}E_{A}+ \simS^{A}_{0,0}E_{B},\label{eq:H1_a}\\
   \boldsymbol{\simH}_{A,0} =&\; \boldsymbol{\simS}^{A}_{A,0}E_{A}+ \boldsymbol{\simS}^{A}_{A,0}E_{B},\label{eq:H1_b}\\
    \boldsymbol{\simH}_{0,A} =&\; \boldsymbol{\simH}^{A}_{0,A}+ \boldsymbol{\simS}^{A}_{0,A}E_{B},\label{eq:H1_c}\\
    \boldsymbol{\simH}_{A,A} =&\; \boldsymbol{\simH}^{A}_{A,A}+ \boldsymbol{\simS}^{A}_{A,A}E_{B}. \label{eq:H1_d}
\end{align}
derived in Appendix \ref{sec:extensivity}. The superscript $A$ denotes that the matrix element only refers to quantities of system $A$; e.g., $ \simS^{A}_{0,0} = \Tbraket{\mathrm{R}_A}{\simP_A}{\mathrm{R}_A}$.
We may recast the generalized eigenvalue equation of $\{\boldsymbol{\simH}_1, \boldsymbol{\simS}_1\}$ as
\begin{align}
\begin{split}
  &\det \Bigg(\begin{pmatrix}
      \simS^A_{0} E_{A} & 
     \boldsymbol{\simH}^A_{0,A}\\
    \boldsymbol{\simS}^A_{A, 0}E_A & \boldsymbol{\simH}^A_{A,A}
  \end{pmatrix} - (E-E_{B})\begin{pmatrix}
      \simS^{A}_{0,0} & 
    \boldsymbol{\simS}^{A}_{0,A}\\
    \boldsymbol{\simS}^{A}_{A,0} & \boldsymbol{\simS}^{A}_{A,A}
  \end{pmatrix} \Bigg) = 0.
  \end{split}
\end{align}
This is equivalent to the EOM-ECC equations for system $A$, and hence, the eigenvalues $E_A + \omega_A = E - E_B$ are the energies of system $A$. The total energy is $E = E_A + E_B + \omega_A$. Therefore, we can conclude that the spectrum of $\boldsymbol\simH_1$ contains ground state energy and the excited state energies corresponding to an excitation in system $A$.

We now proceed with $\boldsymbol{\simH}_3$ and $\boldsymbol{\simS}_3$, and we will show that the excited state energies of system $B$ can be found in this generalized eigenvalue problem.
In a separate calculation on system $B$, we identify $\boldsymbol{r}^B$ as a right eigenvector of the Jacobian matrix with eigenvalue $\omega_B$:
\begin{align}
  \boldsymbol{J}_{B,B}\boldsymbol{r}^B = \omega_B \boldsymbol{r}^B.
\end{align}
We now want to demonstrate that the vector 
\begin{align}
    \boldsymbol{X} = \begin{pmatrix}
        \boldsymbol{r}^B \\
        \boldsymbol{0}
    \end{pmatrix}
\end{align}
is a generalized eigenvector of $\{\boldsymbol{\simH}_3,\boldsymbol{\simS}_3\}$. We have
\begin{align}
    \boldsymbol{\simH}_3\boldsymbol{X} &= 
    \begin{pmatrix}
   \boldsymbol{\simH}_{B,B} & \boldsymbol{\simH}_{B,AB}\\
   \boldsymbol{\simH}_{AB,B} & \boldsymbol{\simH}_{AB,AB}\\
    \end{pmatrix}
     \begin{pmatrix}
     \boldsymbol{r}^B \\
     \boldsymbol{0}
     \end{pmatrix}
    = 
    \begin{pmatrix}
      \boldsymbol{\simH}_{B,B} \boldsymbol{r}^B\\
      \boldsymbol{\simH}_{AB,B} \boldsymbol{r}^B
    \end{pmatrix} 
    = 
    (E_A + E_B + \omega_B)
    \begin{pmatrix}
    \simS^A_{0,0}\boldsymbol{r}^B\\
    \boldsymbol{\simS}^A_{A,0}\otimes\boldsymbol{r}^B\\
    \end{pmatrix}\label{eq:H3}\\
    \boldsymbol{\simS}_3\boldsymbol{X} &= 
    \begin{pmatrix}
   \boldsymbol{\simS}_{B,B} & \boldsymbol{\simS}_{B,AB}\\
   \boldsymbol{\simS}_{AB,B} & \boldsymbol{\simS}_{AB,AB}\\
    \end{pmatrix}
    \begin{pmatrix}
     \boldsymbol{r}^B \\
     \boldsymbol{0}
     \end{pmatrix}
    = 
    \begin{pmatrix}
      \boldsymbol{\simS}_{B,B} \boldsymbol{r}^B\\
      \boldsymbol{\simS}_{AB,B} \boldsymbol{r}^B
    \end{pmatrix} = 
    \begin{pmatrix}
    \simS^A_{0,0}\boldsymbol{r}^B\\
    \boldsymbol{\simS}^A_{A,0}\otimes\boldsymbol{r}^B\\
    \end{pmatrix},
\end{align}
where a detailed derivation is given in Appendix \ref{sec:extensivity}.
Hence, we have $\boldsymbol{\simH_3}\boldsymbol{X} = (E_A + E_B + \omega_B)\boldsymbol{\simS_3}\boldsymbol{X}$, and the generalized eigenvalues correspond to the energy of an excited state in system $B$ (with excitation energy $\omega_B$) and the ground state of system $A$.
With this, we conclude that the EOM-ECC energies are size-extensive.

\end{widetext}

\begin{table*}[hbt]
    \centering
    \caption{The redundant parameters in the ECCSD calculation with a single bath orbital. Sub- and superscripts $*$ denote general occupied or virtual indices that are not $I$ or $A$, respectively.}
    \begin{tabular}{c c c}
    \toprule
    Model & Singles amplitudes & Doubles amplitudes \\
    \midrule
    $\PQa\exp(T)\ketR$ & $t^{A}_{I}$ & $t^{*A}_{*I}$, $t^{AA}_{**}$, $t^{**}_{II}$, $t^{AA}_{II}$\\
    $\PP\exp(T)\ketR$ & $t^*_I$, $t^{A}_{I}$ & $t^{**}_{II}$, $t_{II}^{A*}$, $t^{**}_{I*}$, $t^{*A}_{I*}$, $t^{A*}_{I*}$, $t^{AA}_{I*}$, $t^{AA}_{II}$\\
    $\QQ\exp(T)\ketR$ & $t^{A}_*$, $t^{A}_{I}$ & $t^{AA}_{**}$, $t^{AA}_{I*}$, $t^{A*}_{**}$, $t^{A*}_{*I}$, $t^{A*}_{I*}$, $t^{A*}_{II}$, $t^{AA}_{II}$ \\
    \bottomrule
    \end{tabular}
    \label{tab:redundant_t_PP_QQ}
\end{table*}

\begin{table}[hbt]
    \centering
    \caption{Geometries of doublet molecular systems}
    \begin{tabular}{c c}
    \toprule
    Molecule & Geometry or reference \\
    \midrule
     H$_2$O$^+$  & Ref.~\citenum{olsen1996full} \\
     OH          & Ref.~\citenum{stanton1994}   \\
     CH          & Ref.~\citenum{stahl2022quantifying} \\
     CN          & Ref.~\citenum{stahl2022quantifying}  \\
     HF$^+$      & $\SI{0.917}{\angstrom}$ \\
     N$_3$       & Ref.~\citenum{stahl2022quantifying} \\
     NO$_2$      & Ref.~\citenum{stanton1994}  \\
     \bottomrule
    \end{tabular}
    \label{tab:geometries}
\end{table}

\section{Entanglement coupled cluster theory for doublet and singlet systems}
\subsection{Doublet states}
To determine doublet systems in ECC, we can use either $\P  = \PQa$  or $\P  = \PQb$ in the procedure outlined in Section \ref{sec:ECC} to obtain  $S_{\mathrm{m}z} = \pm\frac{1}{2}$ for the molecule. 
We choose $S_{\mathrm{m}z} = \frac{1}{2}$, and therefore use the projection $\P  = \PQa$. 
Upon application of the projector to the coupled cluster state, some parameters are redundant and must be eliminated. In Table \ref{tab:redundant_t_PP_QQ}, we list the parameters that are explicitly removed.

 In an ECC calculation for a doublet system of $N_\mathrm{m}$ electrons,  we can use the ROHF orbitals of the target $N_\mathrm{m}$-electron system or the RHF orbitals from an $(N_\mathrm{m} + 1)$-electron calculation.
If we use ROHF orbitals, the singly occupied orbital is taken to be active and is mixed with the bath orbital. If, on the other hand, we use RHF orbitals, the highest occupied molecular orbital (HOMO) is chosen to be active. 

For the projected spin of the bath, we can show that
\begin{align}
    S_{\mathrm{b}z} \PQa = -\frac{1}{2} \PQa,
\end{align}
and it follows that 
\begin{align}
    S_{\mathrm{b}z} \PQa \ket{\Psi} = -\frac{1}{2}\PQa \ket{\Psi},
\end{align}
as long as $\PQa \ket{\Psi} \neq 0$. For the spin projection of the molecule, we have (see Appendix \ref{sec:spin} for a detailed derivation)
\begin{align}
    S_{\mathrm{m}z} \PQa \ket{\Psi} = -S_{\mathrm{b}z} \PQa \ket{\Psi} = \frac{1}{2}\PQa \ket{\Psi}.
\end{align}
For the total spin, we use the relation
\begin{align}
\vec{S}^2_{\mathrm{b}} = S_{\mathrm{b}z}(S_{\mathrm{b}z}-1) + S_{\mathrm{b}+}S_{\mathrm{b}-}
\end{align}
and one may show that
\begin{align}
S_{\mathrm{b}+}S_{\mathrm{b}-}\PQa  = 0,
\end{align}
which implies for the squared spin
\begin{align}
\begin{split}
\vec{S}^2_{\mathrm{b}}\PQa \ket{\Psi} &= S_{\mathrm{b}z}(S_{\mathrm{b}z}-1)\PQa \ket{\Psi} \\
&= -\frac{1}{2}\Bigl(-\frac{1}{2}-1\Bigr)\PQa \ket{\Psi} \\
&=  \frac{1}{2}\Bigl(\frac{1}{2}+1\Bigr)\PQa \ket{\Psi}.
\end{split}
\end{align}
With this, we have established that $S_\mathrm{b} = \frac{1}{2}$. From the addition theorem (the Clebsch-Gordan series), we conclude that $S_\mathrm{m} = \frac{1}{2}$, such that $|S_\mathrm{m} - S_\mathrm{b}| = 0$.
The state $\P \exp(T)\ketR = \PQa \exp(T)\ketR$ is, therefore, a spin-pure state for the molecular doublet system. 
\begin{figure*}
    \centering
    \includegraphics[width=\linewidth]{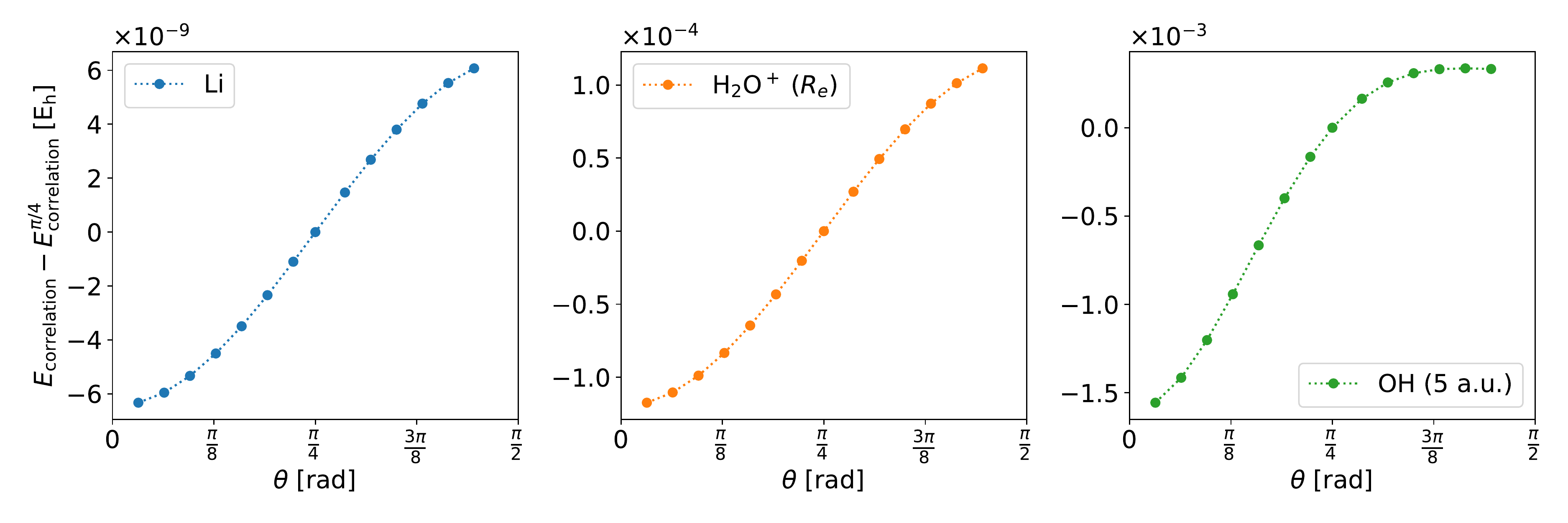}
    \caption{$\theta$-dependence of ECCSD correlation energies ($E_{\mathrm{correlation}}$) for Li (left), H$_2$O$^+$ (middle) and OH at $5$ a.u. bond length (right) using the cc-pVDZ basis. The correlation energies are presented as the difference to that obtained with $\theta=\pi/4$ ($E_{\mathrm{correlation}}^{\pi/4}$). Note that the correlation energy differences are given in scientific notation, and that the differences are smaller than the errors compared to FCI (see Table \ref{tab:H2O_Li_comparison_to_fci}).}
    \label{fig:theta-dependence}
\end{figure*}
\begin{figure*}
    \centering
    \includegraphics[width=\textwidth]{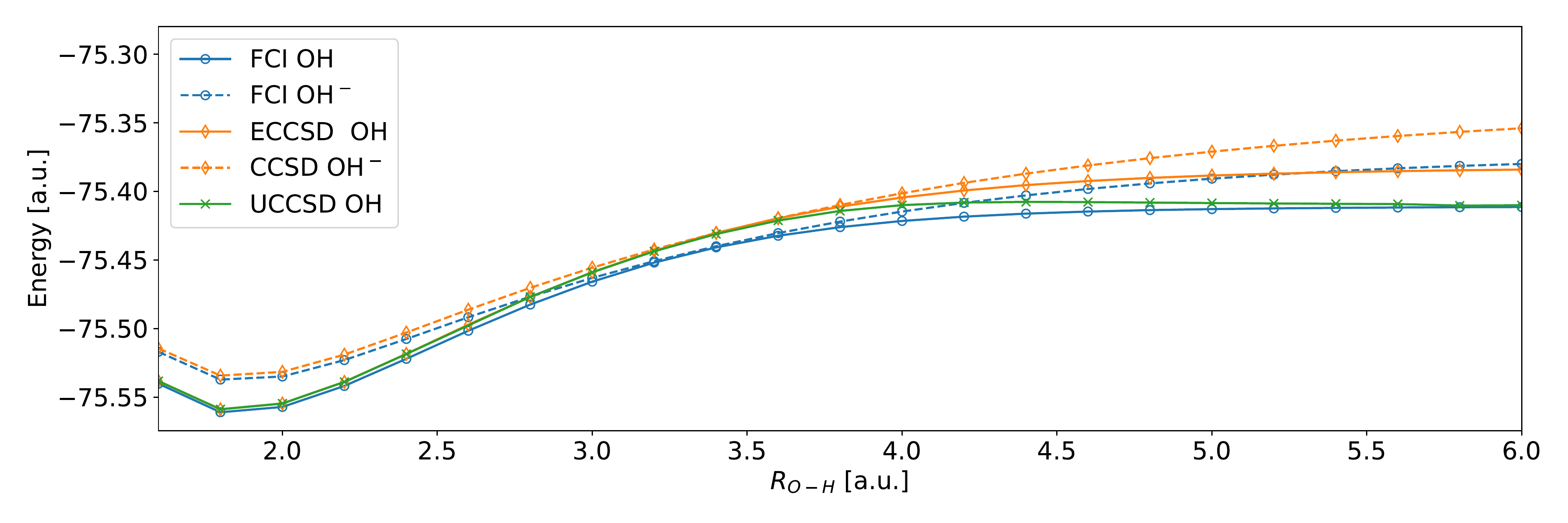}
    \includegraphics[width=\textwidth]{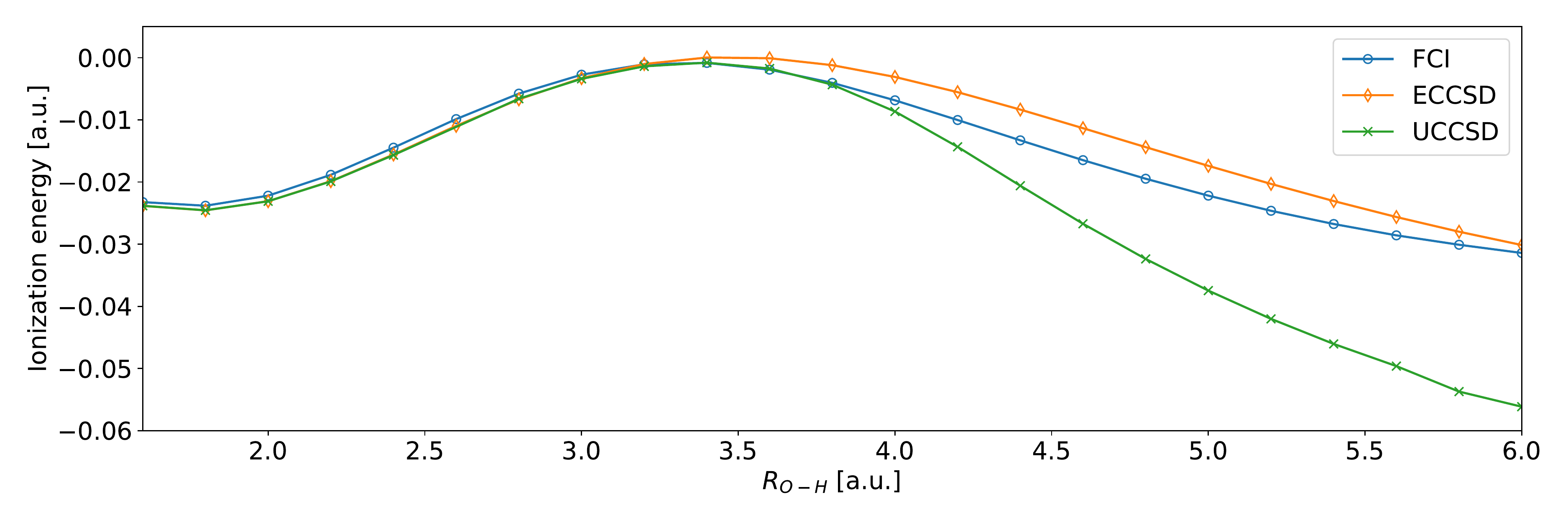}
    \caption{\revS{Top: dissociation curves for OH/cc-pVDZ (solid lines) and OH$^-$/cc-pVDZ (stipled lines) with FCI, CCSD, ECCSD and UCCSD. Bottom: ionization energies with FCI, UCCSD and ECCSD for OH$^-$/cc-pVDZ. Closed-shell CCSD calculation is performed with spin adaptation.}}
    \label{fig:dissociation}
\end{figure*}

\subsection{Singlet states}
\revS{All the projection operators in Eqs. \eqref{eq:PP} -- \eqref{eq:doublet-spin-projector} are orthogonal. Consequently, the corresponding ECC states $\P \exp(T)\ketR$ are orthogonal and since all the projectors commute with $H$, the different states are also non-interacting with $H$.}

\revS{There are two projection operators that result in a singlet state for both the molecule and the bath, see Eqs. \eqref{eq:PP} and \eqref{eq:QQ}. 
For $\P = \QQ$, there are 0 electrons in the bath after projection, that is, $N_\mathrm{m} = N_e$. For $\P = \PP$, there are 2 electrons in the bath after projection, that is, $N_\mathrm{m} = N_e - 2$.}
\revS{With} both \revS{of} these \revS{projectors}, the standard CCSD energy is obtained \revS{with ECCSD. This is because the additional amplitudes, obtained by adding the bath orbital,} are redundant. In Table \ref{tab:redundant_t_PP_QQ}, we show explicitly which amplitudes are redundant in the two cases.


\begin{table*}[hbt]
    \caption{Energies of different doublet systems in the cc-pVDZ basis, computed with entanglement CCSD (ECCSD), unrestricted CCSD (UCCSD), ROHF, and UHF.
    The e$^\mathcal{T}$ program is used for ECCSD and ROHF, PySCF\citep{sun2015libcint, sun2018pyscf, sun2020recent} is used for UCCSD and UHF, and CFOUR is used for SR-CCSD.\cite{matthews2020coupled} All energies are given in Hartree. We also give the spin contamination obtained for the UHF calculation, defined as $\Delta S = \langle S^2\rangle^\mathrm{UHF} - \langle S^2\rangle_{\mathrm{exact}}$.}
    \centering{
    \begin{tabular}{l c c c c c c c}
    \toprule
    System & $E_{\mathrm{ECCSD}}$ (ROHF) & $E_{\mathrm{ECCSD}}$ (RHF) &  $E_{\mathrm{UCCSD}}$ & $E_{\mathrm{SR-CCSD}}$ &
    $E_{\mathrm{ROHF}}$ &
    $E_{\mathrm{UHF}}$ &
    $\Delta S$ [a.u.]\\
    \midrule
    CH              & -38.379549  & -38.380134 & -38.379603  & -38.378857  & -38.268536  & -38.272381  & $7.2\times10^{-3}$\\
    OH              & -75.559364  & -75.559216 & -75.559360  & -75.558918  & -75.390010  & -75.393846  & $4.6\times10^{-3}$\\ 
    H$_2$O$^+$      & -75.804269  & -75.803823 & -75.804287  & -75.729350$^\dagger$  & -75.629494  & -75.634062  & $6.4\times10^{-3}$\\
    CN              & -92.480435  & -92.440992 & -92.480373  & -92.471770  & -92.191440  & -92.211419  & $4.9\times10^{-1}$\\
    HF$^+$          & -99.661159  & -99.660001 & -99.661163  & -99.660569  & -99.494912  & -99.498992  & $3.4\times10^{-3}$\\ 
    N$_3$           & -163.724366 & -163.726293 & -163.726072 & -163.722451 & -163.228642 & -163.256375 & $1.7\times10^{-1}$\\
    NO$_2$          & -204.534261 & -204.513678 & -204.534472 & -204.350988 & -203.957530 & -203.971743 & $4.9\times10^{-2}$\\
    \bottomrule
    \end{tabular}
    \label{tab:doublets}
    }\\
    \raggedright
    $^\dagger$ ROHF reference energy is different from e$^\mathcal{T}$.
\end{table*}

\section{Extension to triplet states}
\revS{An extension to higher values of $S_\mathrm{m}$ with the ECC approach results in a significant increase in complexity of the working equations. An electron bath with more than a single orbital necessitates the use of automated generation of equations and code. In this section, we outline the procedure to obtain triplet states within the ECC framework.} 

\revS{Considering} two active orbitals and two bath orbitals, we can choose to mix the active and bath orbitals in the following way:
\begin{align}
\psi_I &= \phi_{\mathcal{A}_1}\cos\theta - \phi_{\mathcal{B}_1}\sin\theta, \\
\psi_A &=  \phi_{\mathcal{A}_1}\sin\theta + \phi_{\mathcal{B}_1} \cos\theta,\\
\psi_J &= \phi_{\mathcal{A}_2}\cos\theta - \phi_{\mathcal{B}_2}\sin\theta, \\
\psi_B &=  \phi_{\mathcal{A}_2}\sin\theta +  \phi_{\mathcal{B}_2}\cos\theta,
\end{align}
\revS{where we have chosen to use a single mixing angle $\theta$.}
A minimal example of a triplet system has two orbitals. In this case, the corresponding mixed-orbital reference becomes
\begin{equation}
    \ketR = 
    {a}^\dagger_{I\alpha}{a}^\dagger_{I\beta}{a}^\dagger_{J\alpha}{a}^\dagger_{J\beta}\ket{\text{vac}}.\label{eq:triplet_ref}
\end{equation}
\revS{This state can be expressed} in the original MO basis. There are 16 resulting terms\revS{. However,} if we only consider those with two electrons in the \revS{molecule} (and consequently two electrons in the bath)\revS{, we obtain}
\begin{align}
\begin{split}
    \ketR = C&\Big(
    c^\dagger_{\mathcal{A}_1\alpha} c^\dagger_{\mathcal{A}_1\beta}c^\dagger_{\mathcal{B}_2\alpha} c^\dagger_{\mathcal{B}_2\beta}
    + c^\dagger_{\mathcal{B}_1\alpha} c^\dagger_{\mathcal{B}_1\beta}c^\dagger_{\mathcal{A}_2\alpha} c^\dagger_{\mathcal{A}_2\beta}\\
    &+ c^\dagger_{\mathcal{A}_1\alpha} c^\dagger_{\mathcal{B}_1\beta}c^\dagger_{\mathcal{A}_2\alpha} c^\dagger_{\mathcal{B}_2\beta}
    + c^\dagger_{\mathcal{A}_1\beta}c^\dagger_{\mathcal{B}_1\alpha}c^\dagger_{\mathcal{A}_2\beta} c^\dagger_{\mathcal{B}_2\alpha} \\
    &- c^\dagger_{\mathcal{A}_1\alpha} c^\dagger_{\mathcal{B}_1\beta}c^\dagger_{\mathcal{A}_2\beta}c^\dagger_{\mathcal{B}_2\alpha} 
    - c^\dagger_{\mathcal{A}_1\beta}c^\dagger_{\mathcal{B}_1\alpha} c^\dagger_{\mathcal{A}_2\alpha} c^\dagger_{\mathcal{B}_2\beta}\Big)\ket{\text{vac}}\\
    +&\text{ (terms with $N_m\neq2$})
\end{split}
\end{align}
with $C = \cos^2\theta\sin^2\theta$.
The first two terms correspond to \revS{closed-shell singlets for the molecule}. The next two terms are high-spin triplet configurations. The last two terms correspond to a linear combination of \revS{a} low-spin triplet and \revS{an} open-shell singlet configuration.

To extract a high-spin triplet configuration for the \revS{molecule}, we may use the projection operators 
\begin{align}
\begin{split}
    \P_{1} = P_{1}^{\alpha}P_{2}^{\alpha}Q_{1}^{\beta}Q_{2}^{\beta}\\
    \P_{\text{-}1} = P_{1}^{\beta}P_{2}^{\beta}Q_{1}^{\alpha}Q_{2}^{\alpha},
\end{split} 
\end{align} 
ensuring either two $\alpha$-electrons or $\beta$-electrons in the system. 

\section{Results of ECCS and ECCSD for doublet systems\label{sec:results}}
The ECCS ground and excited states, and the ECCSD ground state have been implemented in a development version of the  e$^\mathcal{T}$ program.\citep{folkestad2020t}
This implementation is not optimized and does not have the correct $\mathcal{O}(N^6)$ scaling. In the future, we plan to use automatic code generation to implement the optimal scaling algorithm.

 In the following, we use e$^\mathcal{T}$ for the reported EOM-CCSD and EOM-CC3 calculations,\citep{folkestad2020t,paul2020new} in addition to the ECC calculations.
Time-dependent ROHF (TD-ROHF) and FCI calculations are carried out with the Dalton program,\citep{dalton} and unrestricted CCSD (UCCSD) and
 spin-restricted CCSD (SR-CCSD) calculations are performed with PySCF\citep{sun2015libcint, sun2018pyscf, sun2020recent} and CFOUR,\cite{matthews2020coupled} respectively.

We present calculations for several small doublet molecular systems, and in Table \ref{tab:geometries}, we give {their} geometries. We use the cc-pVDZ basis set throughout. Unless otherwise stated, the mixing angle $\theta$ equals $\frac{\pi}{4}$. 

In Table \ref{tab:doublets}, we present ECCSD calculations on a selection of doublet systems at equilibrium geometries. We use both ROHF and RHF ($N_e = N_\mathrm{m} + 1$) orbitals and compare to  UCCSD with a UHF reference. We also give the ROHF and UHF energies for these systems, and report the UHF spin contamination, defined as {$\Delta S = \langle \vec{S}^2\rangle^\mathrm{UHF} - \langle \vec{S}^2\rangle_{\mathrm{exact}}$. }

The results show that the ECCSD and UCCSD energies are very close, with \sarai{differences} no larger than $\si{\milli\hartree}$, even when there is significant spin contamination in the UHF reference. Hence, we confirm the well-known result that UCCSD successfully removes the spin contamination of the reference through the inclusion of dynamical correlation for systems around their equilibrium geometries. 
\sarai{The differences to SR-CCSD are of the same magnitude, as long as the ROHF reference is the same.}

We may also conclude that RHF orbitals can be used without significant loss of accuracy in ECCSD calculations. \sarai{That is,} the ECCSD model for doublet systems can be used without an implementation of ROHF. 

In Table \ref{tab:H2O_Li_comparison_to_fci}, we compare ECCSD to FCI. The errors are comparable to those of the spin-adapted closed-shell theory.\citep{olsen1996full}

\begin{table}
    \centering
    \caption{Comparison of the ECCSD energies of Li and H$_2$O$^+$ in the cc-pVDZ basis obtained with $\theta=\pi/4$ and the FCI energies. Errors ($\Delta _\mathrm{ECCSD}= E_{\mathrm{ECCSD}} - E_{\mathrm{FCI}}$) are also given. Energies are given in Hartree.}
    \begin{tabular}{l c c c}
    \toprule
    System & $E_{\mathrm{ECCSD}}$ &
    $E_{\mathrm{FCI}}$   & 
    $\Delta_\mathrm{ECCSD}$\\
    \midrule
    Li                  & $-7.432637$  &  $-7.432637$ & $2.3\times10^{-7}$\\
    H$_2$O$^+$  & $-75.804269$ & $-75.806892$ & $2.6\times10^{-3}$\\
    OH ($R_e$)          & $-75.559364$ & $-75.561568$ & $2.2\times10^{-3}$\\ 
    OH ($5$ a.u.)       & $-75.388396$ & $-75.412894$ & $2.4\times10^{-2}$\\ 
    HF$^+$              & $-99.661159$ & $-99.662959$ & $1.8\times10^{-3}$\\ 
    \bottomrule
    \end{tabular}
    \label{tab:H2O_Li_comparison_to_fci}
\end{table}

\subsection{Dependence on the mixing angle}

The mixing parameter $\theta$ in Eqs. \eqref{eq:occ_mixed}, \eqref{eq:vir_mixed} and \eqref{eq:theta_transform} enters throughout the ECC equations; even when RHF orbitals are used. For ECCS, there is no theta dependence on the ground and excited state energies. For ECCSD, however, there is a small energy dependence on $\theta$, except for systems with only two electrons. In Fig. \ref{fig:theta-dependence}, we have plotted the variation in the energy as a function of $\theta$ for Li, H$_2$O$^+$, and OH ($5$ a.u.) relative to the energy obtained with $\theta=\frac{\pi}{4}$. Note that we use a scientific scale, with exponents given in the top left corner. Comparing to Table \ref{tab:H2O_Li_comparison_to_fci}, we see that the variation in the energy with $\theta$ is at least an order of magnitude smaller than the error to FCI for these systems. For Li and H$_2$O$^+$ at equilibrium geometry, we see that the variation is symmetric about $\theta=\frac{\pi}{4}$, but this is not the case for OH when the OH bond is stretched.
\begin{table*}[]
    \centering
    \caption{Ionization potential in $\si{\eV}$, obtained with different coupled cluster methods and FCI. For ECCSD, UCCSD and FCI, the ionization potential is obtained by differences in total energies. FCI calculations are performed using the Dalton program,\citep{dalton} the UCCSD calculations are performed with PySCF,\citep{sun2015libcint, sun2018pyscf, sun2020recent} and all remaining calculations are performed with e$^\mathcal{T}$.\citep{folkestad2020t,paul2020new}}
    \begin{tabular}{l c c c c c}
    \toprule
    System & 
    $E_{\mathrm{ECCSD}}$  &
    $E_{\mathrm{UCCSD}}$ &
    $E_{\mathrm{EOM-CCSD}}$ & 
    $E_{\mathrm{EOM-CC3}}$ & 
    $E_{\mathrm{FCI}}$ \\
    \midrule
    HF & 15.4286 & 15.4286 & 15.2030 & 15.4717 & 15.4459 \\
    H$_2$O  & 11.8053 & 11.8051 & 11.6673 & 11.8355  & 11.8358\\
    OH$^-$  & -0.6657 & -0.6654 & -0.7613 & -0.6354 & -0.6448 \\
    \bottomrule
    \end{tabular}
    \label{tab:comparison_to_eom_ip}
\end{table*}
\subsection{Ionization energies}
\revS{In Fig. \ref{fig:dissociation}, we present the dissociation curves of OH/cc-pVDZ and OH$^-$/cc-pVDZ calculated with CCSD (spin-adapted closed-shell), ECCSD, UCCSD, and FCI (top panel), and the corresponding ionization energies (bottom panel). The ionization energies are calculated as $E_{\text{IP}} = E_{\text{OH}} - E_{\text{OH}^-}$. Since the open-shell (ionized) system has lower energy, the IP is negative.}
\revS{The ECCSD dissociation curve is similar in shape to the FCI dissociation for OH.
For UCCSD, on the other hand, the dissociation curve differs in shape from the FCI curve: it displays a maximum around ${4}$ a.u.
At $6$ a.u., UCCSD displays significantly smaller errors compared to FCI ($1.3\times10^{-3}\si{\hartree}$) than ECCSD. As a result, UCCSD offers a less balanced description for ionization energies at large bond lengths.} \sarai{At intermediate bond lengths, UCCSD ionization energies are closer to FCI.}

In Table \ref{tab:comparison_to_eom_ip}, we compare ionization energies obtained with ECCSD, UCCSD, EOM-CCSD, EOM-CC3, and FCI. For ECCSD, UCCSD, and FCI the ionization energies are given by the difference between the ionized doublet state and the closed-shell initial state. For EOM-CCSD and EOM-CC3, the ionization energies are obtained by including a non-interacting orbital in the standard EOM-CC calculation. This is implemented through a projection, equivalent to the implementation of core excited states from the core-valence-separation (CVS) approach as described in Refs. \citenum{coriani2015communication} and \citenum{coriani2016erratum}. Again, we see that ECCSD and UCCSD yield similar results around equilibrium geometries. The accuracy is comparable to the EOM-CC3 ionization energies for these systems; the errors are around $10^{-2}$ $\si{\eV}$. The quality of EOM-CCSD ionization energies is lower, because one index in the EOM vector $\boldsymbol{R}$ is always restricted to the non-interacting bath orbital.

\subsection{Excitation energies with EOM-ECCS}

\begin{table}[htb]
    \centering
    \caption{The first 9 excitation energies $\omega_i$ [eV] calculated for H$_2$O$^+$ in the cc-pVDZ basis with both the presented EOM-ECCS theory and the Tamm-Dancoff approximation (TDA).}
    \label{tab:ROHF_EOM-CCS_vs_TDA}
    \begin{tabular}{c c c c} 
    \toprule
    $\omega_i$  &        TDA               &     EOM-ECCS            & Difference    \\
    \midrule
    $\omega_1$  &  $ \phantom{0}2.0735  $  &  $ \phantom{0}2.0735  $ & ${1.3\times10^{-9}}$ \\
    $\omega_2$  &  $ \phantom{0}6.9716  $  &  $ \phantom{0}6.9716  $ & ${5.4\times10^{-8}}$  \\
    $\omega_3$  &  $ 15.6874 $             &  $ 15.6874 $            & ${6.3\times10^{-8}}$  \\
    $\omega_4$  &  $ 15.7200 $             &  $ 15.7200 $            & ${2.6\times10^{-8}}$ \\
    $\omega_5$  &  $ 17.5132 $             &  $ 17.5132 $            & ${7.3\times10^{-8}}$  \\
    $\omega_6$  &  $ 17.7718 $             &  $ 17.7718 $            & ${3.5\times10^{-8}}$  \\
    $\omega_7$  &  $ 19.2738 $             &  $ 19.2738 $            & ${3.8\times10^{-8}}$  \\
    $\omega_8$  &  $ 22.6943 $             &  $ 22.6943 $            & ${1.4\times10^{-8}}$ \\
    $\omega_9$  &  $ 22.7759 $             &  $ 22.7759 $            & ${1.3\times10^{-7}}$  \\
    \bottomrule 
    \end{tabular}
\end{table}

\begin{table}[]
    \centering
    \caption{Ground state energies $E_n$ [\si{\hartree}] and the first three excitation energies, $\omega$ [eV], calculated for H$_2$O$^+$ from ROHF orbitals and $n$ non-interacting He atoms (placed \SI{200}{\angstrom} apart) with the cc-pVDZ basis set. The RHF energy of a single isolated Helium atom was calculated to be $E_\text{He}=\SI{-2.85516}{\hartree}$. Differences in excitation energies are of order $10^{-12}$, which corresponds to the convergence threshold.}
    \begin{tabular}{c c c c c c c c}
    \toprule
    &$n$ & $E_n$ & $E_n - E_0 - nE_\text{He}$ & $\omega_1$ & $\omega_2$ & $\omega_3$\\
    \midrule
    \multirow{5}{*}{ROHF}
    &0 & $-75.6295$ &       0                             & $2.0742$ & $6.9706$ & $15.6876$ \\
    &1 & $-78.4847$ & ${-8.0\times 10^{-12}} $            & $2.0742$ & $6.9706$ & $15.6876$ \\
    &2 & $-81.3398$ & ${-8.0\times 10^{-12}}$             & $2.0742$ & $6.9706$ & $15.6876$ \\
    &3 & $-84.1950$ & ${-8.0\times 10^{-12}} $ & $2.0742$ & $6.9706$ & $15.6876$ \\
    &4 & $-87.0501$ & ${-7.0\times 10^{-12}} $ & $2.0742$ & $6.9706$ & $15.6876$ \\
    \midrule
    \multirow{5}{*}{RHF}
    &0 & $-76.0240$ & 0                                  & $1.9469$ & $6.6466$ & $16.4827$ \\
    &1 & $-78.8792$ & ${-4.0\times 10^{-15}} $           & $1.9469$ & $6.6466$ & $16.4827$ \\
    &2 & $-81.7344$ & ${-8.0\times 10^{-15}} $           & $1.9469$ & $6.6466$ & $16.4827$ \\
    &3 & $-84.5895$ & ${\phantom{-}1.0\times 10^{-12}} $ & $1.9469$ & $6.6466$ & $16.4827$ \\
    &4 & $-87.4447$ & ${\phantom{-}1.0\times 10^{-12}} $ & $1.9469$ & $6.6466$ & $16.4827$ \\
    \bottomrule
    \end{tabular}
    \label{tab:size_extensivity}
\end{table}
At the ECCS level of theory, we have implemented both the ground and the excited state equations. With ROHF orbitals, the ground state energy equals the ROHF energy, and the excitation energies equal those obtained with the Tamm-Dancoff approximation (TDA) in TD-ROHF. 
See Table \ref{tab:ROHF_EOM-CCS_vs_TDA}, where we compare to TDA-TD-ROHF excitation energies. 

In Table \ref{tab:size_extensivity}, we demonstrate the size-extensivity of the ECCS approach when non-interacting closed-shell subsystems are added to the calculation. We consider H$_2$O$^+$ + $n$He where the He atoms are placed $\SI{200}{\angstrom}$ from the H$_2$O$^+$ molecule. Both the ground state energy and excitation energies show the correct scaling properties. The ground state energy is size-extensive, and the excitation energies are size-intensive. Here, we present results using both ROHF and RHF ($N_e = N_{\mathrm{m}} + 1$) orbitals. 
With ROHF orbitals, the ECC $t$-amplitudes are all $0$, and the equations converge in one iteration. This is exactly equivalent to CCS with RHF orbitals, and due to the Brillouin theorem.
With RHF orbitals, however, the $t$-amplitudes are different from $0$, and the ECCS equations must be solved iteratively.
As expected, the ECCS results depend more strongly on the orbitals than in ECCSD. However, the size-extensivity properties are demonstrated for both choices of orbitals.


\section{Summary and concluding remarks}
In this paper, we have introduced the entanglement coupled cluster \revS{(ECC)} approach for a spin-adapted treatment of open-shell systems. \revS{We have demonstrated how this approach can be used to obtain ground and excited state energies and that the energies are size-extensive.}

In the ECC approach, the orbitals of the system of interest are mixed with the orbitals of a non-interacting bath. A closed-shell reference is constructed in the mixed orbital basis and is used in the exponential parametrization of coupled cluster theory. Since the reference determinant is closed-shell, the closed-shell spin-adapted formulation of coupled cluster theory can be used. The complications of defining a spin-free operator for an open-shell reference are avoided and the BCH expansion of the similarity-transformed Hamiltonian truncates after only four nested commutators with the cluster operator.
In order to obtain the actual system of interest, a projection operator is applied to the coupled cluster state. 

A pilot implementation is presented for the ECCSD doublet ground state, in addition to the ground and excited states with ECCS. Proof-of-concept calculations demonstrate that the error of the ECCSD energy is comparable to that of spin-adapted closed-shell CCSD. For OH and OH$^-$, we have demonstrated that this holds at all bond lengths. ECCS energies equal the ROHF energies, when ROHF orbitals are used, and for the excited states, ECCS yields Tamm-Dancoff TD-ROHF energies, analogous to the relation between CCS and RHF.

\section{Acknowledgments}
This work has received funding from the European Research Council (ERC) under the European Union’s Horizon 2020 Research and Innovation Programme (grant agreement No. 101020016). S.D.F acknowledges funding from ``Fondet til professor Leif Tronstads minne" and S.D.F and H.K both acknowledge funding from the Research Council of Norway through FRINATEK project 275506.
We acknowledge computing resources through UNINETT Sigma2 -- the National Infrastructure for High Performance Computing and Data Storage in Norway, through project number NN2962k. 

\appendix

\section{Size-extensivity\label{sec:extensivity}}
Since the systems $A$ and $B$ are non-interacting, we have $H = H_A + H_B$ and  $T = T_A + T_B$. Since system $B$ is closed-shell, $\mathcal{P} = \mathcal{P}_A$ commutes with $T_B$ and $H_B$. Hence, we have
\begin{align}
    \simP = \exp(-T_A)\mathcal{P}\exp(T_A) =\mathcal{P}_A^{T_A}. 
\end{align}
The matrix elements in Eq. \eqref{eq:H1} are evaluated below.
For $\bar{H}_{0,0}$, we have
\begin{align}
\begin{split}
    \bar H_{0,0} =& \Tbraket{\mathrm{R}_A}{\mathcal{P}^{T_A}_A H^{T_A}_A}{\mathrm{R}_A}\Dbraket{\mathrm{R}_B}{\mathrm{R}_B} \\
    &+ \Tbraket{\mathrm{R}_A}{\mathcal{P}^{T_A}_A }{\mathrm{R}_A}\Tbraket{\mathrm{R}_B}{H^{T_B}_B}{\mathrm{R}_B}\\
    =& \Tbraket{\mathrm{R}_A}{\mathcal{P}^{T_A}_A H^{T_A}_A}{\mathrm{R}_A} \\
    &+ \Tbraket{\mathrm{R}_A}{\mathcal{P}^{T_A}_A }{\mathrm{R}_A}\Tbraket{\mathrm{R}_B}{H^{T_B}_B}{\mathrm{R}_B}  \\
    =&\bar S^{A}_{0,0}E_{A}+ \bar S^{A}_{0,0}E_{B},
\end{split}
\end{align}
where we have used the definitions of the ground state energy of systems $A$ and $B$:
\begin{align}
\begin{split}
        E_A S^A_{0,0} &= {E_A\Tbraket{\mathrm{R}_A}{\mathcal{P}^{T_A}_A}{\mathrm{R}_A}}\\
        &=\Tbraket{\mathrm{R}_A}{\mathcal{P}^{T_A}_A H^{T_A}_A}{\mathrm{R}_A}
\end{split}\\
    E_B &= \Tbraket{\mathrm{R}_B}{H^{T_B}_B}{\mathrm{R}_B}
\end{align}
We have also used $\Dbraket{\mathrm{R}_B}{\mathrm{R}_B} = 1$.
For $\boldsymbol{\bar{H}}_{A,0}$, we have
\begin{align}
\begin{split}
    \boldsymbol{\bar{H}}_{A,0} =& \Tbraket{\boldsymbol{\mu}_A}{\mathcal{P}_A^{T_A} H^{T_A}_A}{\mathrm{R}_A}\Dbraket{\mathrm{R}_B}{\mathrm{R}_B} \\
    &+ \Tbraket{\boldsymbol{\mu}_A}{\mathcal{P}_A^{T_A} }{\mathrm{R}_A}\Tbraket{\mathrm{R}_B}{H^{T_B}_B}{\mathrm{R}_B}\\
    =& \Tbraket{\boldsymbol{\mu}_A}{\mathcal{P}_A^{T_A} H^{T_A}_A}{\mathrm{R}_A} \\
    &+ \Tbraket{\boldsymbol{\mu}_A}{\mathcal{P}_A^{T_A} }{\mathrm{R}_A}\Tbraket{\mathrm{R}_B}{H^{T_B}_B}{\mathrm{R}_B}  \\
     =& \boldsymbol{\bar S}^{A}_{A,0}E_{A}+ \boldsymbol{\bar S}^{A}_{A,0}E_{B},
\end{split}
\end{align}
where, in the last line, we have used the ground state ECC equations for subsystem $A$:
\begin{align}
    \Tbraket{\boldsymbol{\mu}_A}{\mathcal{P}_A^{T_A} H^{T_A}_A}{\mathrm{R}_A}  &= \Tbraket{\boldsymbol{\mu}_A}{\mathcal{P}_A^{T_A}}{\mathrm{R}_A}E_A\\
    &= \boldsymbol{\bar S}^{A}_{A,0}E_{A}.
\end{align}

For $\boldsymbol{\bar{H}}_{0,A}$, we have
\begin{align}
\begin{split}
    \boldsymbol{\bar{H}}_{0,A} =& \Tbraket{\mathrm{R}_A}{\mathcal{P}^{T_A}_A H^{T_A}_A}{\boldsymbol{\mu}_A}\Dbraket{\mathrm{R}_B}{\mathrm{R}_B} \\
    &+ \Tbraket{\mathrm{R}_A}{\mathcal{P}^{T_A}_A }{\boldsymbol{\mu}_A}\Tbraket{\mathrm{R}_B}{H^{T_B}_B}{\mathrm{R}_B}\\
    =& \Tbraket{\mathrm{R}_A}{\mathcal{P}^{T_A}_A H^{T_A}_A}{\boldsymbol{\mu}_A} \\
    &+ \Tbraket{\mathrm{R}_A}{\mathcal{P}^{T_A}_A }{\boldsymbol{\mu}_A}\Tbraket{\mathrm{R}_B}{H^{T_B}_B}{\mathrm{R}_B}  \\
    =& \boldsymbol{\bar{H}}^{A}_{0,A} + \boldsymbol{\bar S}^{A}_{0,A}E_{B}.
\end{split}
\end{align}
And similarly for $\boldsymbol{\bar{H}}_{A,A}$, we have
\begin{align}
\begin{split}
    \boldsymbol{\bar{H}}_{A,A} =& \Tbraket{\boldsymbol{\mu}_A}{\mathcal{P}^{T_A}_A H^{T_A}_A}{\boldsymbol{\nu}_A}\Dbraket{\mathrm{R}_B}{\mathrm{R}_B} \\
    &+ \Tbraket{\boldsymbol{\mu}_A}{\mathcal{P}^{T_A}_A }{\boldsymbol{\nu}_A}\Tbraket{\mathrm{R}_B}{H^{T_B}_B}{\mathrm{R}_B}\\
    =& \Tbraket{\boldsymbol{\mu}_A}{\mathcal{P}^{T_A}_A H^{T_A}_A}{\boldsymbol{\nu}_A} \\
    &+ \Tbraket{\boldsymbol{\mu}_A}{\mathcal{P}^{T_A}_A }{\boldsymbol{\nu}_A}\Tbraket{\mathrm{R}_B}{H^{T_B}_B}{\mathrm{R}_B}  \\
    =& \boldsymbol{\bar{H}}^{A}_{A,A} + \boldsymbol{\bar S}^{A}_{A,A}E_{B}.
\end{split}
\end{align}

For $\boldsymbol{\bar{H}}_{B,B}$, we have
\begin{align}
\begin{split}
    \boldsymbol{\bar{H}}_{B,B} =& \Tbraket{\boldsymbol{\mu}_B}{ H^{T_B}_B}{\boldsymbol{\nu}_B}\Tbraket{\mathrm{R}_A}{\mathcal{P}_A^{T_A}}{\mathrm{R}_A} \\
    &+ \Dbraket{\boldsymbol{\mu}_B}{\boldsymbol{\nu}_B}\Tbraket{\mathrm{R}_A}{\mathcal{P}_A^{T_A}H^{T_A}_A}{\mathrm{R}_A}\\
    =& \Tbraket{\boldsymbol{\mu}_B}{ H^{T_B}_B}{\boldsymbol{\nu}_B}\bar{S}_{0,0}^{A} \\
    &+ \Dbraket{\boldsymbol{\mu}_B}{\boldsymbol{\nu}_B}\bar{S}_{0,0}^{A}E_A\\
    =& \Tbraket{\boldsymbol{\mu}_B}{ H^{T_B}_B}{\boldsymbol{\nu}_B}\bar{S}_{0,0}^{A} +\bar{S}_{0,0}^{A}E_A{\boldsymbol{I}^B}\\
    =& (\Tbraket{\boldsymbol{\mu}_B}{ H^{T_B}_B}{\boldsymbol{\nu}_B} + E_A{\boldsymbol{I}^B})\bar{S}_{0,0}^{A},
\end{split}
\end{align}
where we have assumed a biorthonormal basis for system $B$.
If we let $\boldsymbol{r}_B$ be an eigenvector of the Jacobian matrix of system $B$ with eigenvalue $\omega_B$, then
\begin{align}
\begin{split}
    \Tbraket{\boldsymbol{\mu}_B}{ H^{T_B}_B}{\boldsymbol{\nu}_B} \boldsymbol{r}_B &= (\boldsymbol{J}_B + E_B{\boldsymbol{I}^B})\boldsymbol{r}_B \\
    &= (\omega_B + E_B)\boldsymbol{r}_B.
\end{split}
\end{align}
With this, we have demonstrated that 
\begin{align}
    \boldsymbol{\bar{H}}_{B,B}  \boldsymbol{r}_B = \bar{S}_{0,0}^{A}(\omega_B + E_B + E_A)\boldsymbol{r}_B.
\end{align}

For $\boldsymbol{\bar{H}}_{AB,B}$, we have
\begin{align}
\begin{split}
    \boldsymbol{\bar{H}}_{AB,B} =& \Tbraket{\boldsymbol{\mu}_A}{\mathcal{P}_A^{T_A}}{\mathrm{R}_A}\otimes\Tbraket{\boldsymbol{\mu}_B}{ H^{T_B}_B}{\boldsymbol{\nu}_B}\\
    &+ \Tbraket{\boldsymbol{\mu}_A}{\mathcal{P}_A^{T_A}H^{T_A}_A}{\mathrm{R}_A}\otimes\Dbraket{\boldsymbol{\mu}_B}{\boldsymbol{\nu}_B}\\
    =& \boldsymbol{\bar{S}}_{A,0}^{A}\otimes\Tbraket{\boldsymbol{\mu}_B}{ H^{T_B}_B}{\boldsymbol{\nu}_B} \\
    &+ E_A\boldsymbol{\bar{S}}_{A,0}^{A}\otimes\Dbraket{\boldsymbol{\mu}_B}{\boldsymbol{\nu}_B}\\
    =& \boldsymbol{\bar{S}}_{A,0}^{A}\otimes\Tbraket{\boldsymbol{\mu}_B}{ H^{T_B}_B}{\boldsymbol{\nu}_B}
    +E_A\boldsymbol{\bar{S}}_{A,0}^{A}\otimes{\boldsymbol{I}^B}\\
    =& \boldsymbol{\bar{S}}_{A,0}^{A}\otimes(\Tbraket{\boldsymbol{\mu}_B}{ H^{T_B}_B}{\boldsymbol{\nu}_B}
    + {\boldsymbol{I}^B}E_A).
\end{split}
\end{align}
This yields
\begin{align}
    \boldsymbol{\bar{H}}_{AB,B}\boldsymbol{r}_B& = \boldsymbol{\bar{S}}_{A,0}^{A}\otimes(\Tbraket{\boldsymbol{\mu}_B}{ H^{T_B}_B}{\boldsymbol{\nu}_B}
    + {\boldsymbol{I}^B}E_A)\boldsymbol{r}_B\\
   & = \boldsymbol{\bar{S}}_{A,0}^{A}\otimes(\omega_B + E_B + E_A)\boldsymbol{r}_B.
\end{align}

\section{Spin properties of the ECC doublet\label{sec:spin}}
From the definition of the $S_z$ operator in second quantization,\cite{helgaker2014} we have
\begin{align}
     S_{\mathrm{b}z} = \frac{1}{2}(c^\dagger_{\mathcal{B}\alpha}c_{\mathcal{B}\alpha} - c^\dagger_{\mathcal{B}\beta}c_{\mathcal{B}\beta}) = \frac{1}{2}(P^{\alpha} - P^{\beta}).
\end{align}
Using the properties of the $P^\sigma$ and $Q^\sigma$ operators, we have
\begin{align}
     S_{\mathrm{b}z}\mathcal{P} &= \frac{1}{2}(P^{\alpha} - P^{\beta})P^{\beta}Q^{\alpha}= -\frac{1}{2}P^{\beta}Q^{\alpha},
\end{align}
where we have used $P^\alpha Q^\alpha = 0$ and ${P^\beta}^2 = P^\beta$.
Hence, 
\begin{align}
    S_{\mathrm{b}z} \PQa\exp{(T)}\ketR  = -\frac{1}{2}\PQa\exp{(T)}\ketR\label{eq:spin_b}.
\end{align}
For the total system, we have
\begin{align}
 S_z \PQa \exp{(T)}\ketR =&(S_{\mathrm{m}z} + S_{\mathrm{b}z}) \PQa \exp{(T)}\ketR \\
=& \PQa (S_{\mathrm{m}z} + S_{\mathrm{b}z})\exp{(T)}\ketR\\
=& \PQa S_{z}\exp{(T)}\ketR = 0,
\end{align}
where we have used $[S_{\mathrm{m}z},\PQa] = 0$ and $ [S_{\mathrm{b}z}, \PQa] = 0$, and that $\exp{(T)}\ketR$ is a singlet.
Therefore, we may write
\begin{align}
    S_{\mathrm{m}z}\PQa \exp{(T)}\ketR + S_{\mathrm{b}z}\PQa \exp{(T)}\ketR = 0\\
\implies S_{\mathrm{m}z}\PQa \exp{(T)}\ketR = -S_{\mathrm{b}z}\PQa \exp{(T)}\ketR
\end{align}
and upon insertion of Eq. \eqref{eq:spin_b}, we obtain
\begin{align}
\begin{split}
 S_{\mathrm{m}z}\PQa& \exp{(T)}\ketR \\
 &= -\Big(-\frac{1}{2}\PQa \exp{(T)}\ketR\Big)\\ 
 &= \frac{1}{2}\PQa \exp{(T)}\ketR,
\end{split}
\end{align}
and the ECC doublet state (using $\mathcal{P} = \PQa$) is an eigenvector of $S_{\mathrm{m}z}$ with eigenvalue $\frac{1}{2}$.

We will now prove that 
\begin{align}
    S_{\mathrm{b}+}S_{\mathrm{b}-}\PQa = 0.
\end{align}
From the definition of the raising and lowering operators in second quantization, \cite{helgaker2014} we have
\begin{align}
\begin{split}
    S_{\mathrm{b}+} = a^\dagger_{\mathcal{B}\alpha}a_{\mathcal{B}\beta}\\
    S_{\mathrm{b}-} = a^\dagger_{\mathcal{B}\beta}a_{\mathcal{B}\alpha},
\end{split}
\end{align}
and we have
\begin{align}
\begin{split}
S_{\mathrm{b}+}S_{\mathrm{b}-} &= a^\dagger_{\mathcal{B}\alpha}a_{\mathcal{B}\beta}a^\dagger_{\mathcal{B}\beta}a_{\mathcal{B}\alpha}\\
&= a^\dagger_{\mathcal{B}\alpha}a_{\mathcal{B}\alpha} - a^\dagger_{\mathcal{B}\alpha}a^\dagger_{\mathcal{B}\beta}a_{\mathcal{B}\beta}a_{\mathcal{B}\alpha}\\
&= a^\dagger_{\mathcal{B}\alpha}a_{\mathcal{B}\alpha} - a^\dagger_{\mathcal{B}\alpha}a_{\mathcal{B}\alpha} a^\dagger_{\mathcal{B}\beta}a_{\mathcal{B}\beta} 
\\
&= a^\dagger_{\mathcal{B}\alpha}a_{\mathcal{B}\alpha}(1-a^\dagger_{\mathcal{B}\beta}a_{\mathcal{B}\beta}) \\
&= P^{\alpha}Q^{\beta}.
\end{split}
\end{align}
Since the projection operators are orthogonal (see Eq. \eqref{eq:projection_prop}), we have
\begin{align}
    S_{\mathrm{b}+}S_{\mathrm{b}-}\PQa  =  P^{\alpha}Q^{\beta}\PQa = 0.
\end{align}

\section{ECCSD and ECC doublet equations\label{sec:equations}}
In ECCSD we have
\begin{align}
    \mathcal{P}\exp(T)\ketR 
\end{align}
with $T = T_1 + T_2$ and for doublet systems $\mathcal{P} = P^{\beta}Q^{\alpha}$.
The ground state ECCSD equations read
\begin{align}
    \Tbraket{\mathrm{R}}{\simP\simH}{\mathrm{R}} &= E \Tbraket{\mathrm{R}}{\simP}{\mathrm{R}}\label{eq:ECCSD_E} \\
    \Tbraket{\mathrm{\mu}}{\simP\simH}{\mathrm{R}} &= E \Tbraket{\mu}{\simP}{\mathrm{R}}\label{eq:ECCSD_t}
\end{align}
where $\{\bra{\mu}\} =\{\bra{\mu_1}\}\cup\{\bra{\mu_2}\} $, and where
\begin{align}
  \bra{\mu_i} = \braR\tau_{\mu_i}^\dagger  
\end{align}
with $\tau_{\mu_i}$ for $i = \{1,2\}$ being the single and double singlet excitation operators that define the operator $T$. 
Here, we have defined $\bar{X} =\exp(-T)X\exp(T)$.
We want to solve the Eqs. \eqref{eq:ECCSD_E} and \eqref{eq:ECCSD_t} using the standard machinery of closed-shell spin-adapted coupled cluster theory. Therefore, we find explicit expressions for $\braR\simP$, $\bra{\mu_1}\simP$, and $\bra{\mu_2}\simP$.

In the following, indices $i, j$ denote occupied orbitals including $\psi_I$ and indices $a, b$ denote virtual orbitals including $\psi_A$.
For $\braR\simP$ we can show that the explicit expression becomes
\begin{align}
     \braR{\simP}
    &=C_1\braR
    +C_2\braR E_{IA}^{\beta}
    +C_3\braR E_{IA}^{\alpha} E_{IA}^{\beta}\label{eq:reference}
\end{align}
where we have defined
\begin{align}
\begin{split}
    C_1 =& D_1^2\Big(1 - (t_{I}^{A})^2  - t^{AA}_{II}\Big) + t_{I}^{A}D_1(2D_2 - 1)
\end{split}\\
    C_2 =& D_1(2D_2 - 1 - 2 D_1 t^{A}_I)\\
    C_3 =& -D_1^2
\end{align}
and
\begin{align}
    D_1 =& \sin\theta\cos\theta\\
    D_2 =& \sts\\
    D_3 =& \cts.
\end{align}

For $\bra{\mu_1}\simP$ we can show that the explicit expression becomes
\begin{align}
   \begin{split}
       \bra{\bar{^a_i}}\simP =&C_{1,ai} \braR + C_{2,i}\braR E_{Ia}^\beta + C_{3,a}\braR E_{iA}^\alpha\\
       &+ C_{4, ai} \braR E_{IA}^\beta + C_5 \bra{\bar{^a_i}}\\
       &+ C_{6,i}\braR E_{IA}^\beta E_{Ia}^\alpha + C_{7,a}\braR E_{IA}^\beta E_{iA}^\alpha\\
       &+C_{8}\bra{\bar{^a_i}}E_{IA}^{\beta} + C_{9}\bra{\bar{^a_i}}E_{IA}^{\beta}E_{IA}^{\alpha}
   \end{split} \label{eq:singles}
\end{align}
where we have introduced the additional coefficients:
\begin{align}
\begin{split}
&C_{1,ai} = D_{5,ai}(1-2D_{4}) \\
    &\phantom{C_{1,ai} =}+ {D_1(t^{aA}_{II}D_2 - t^{AA}_{iI}D_3)}
\end{split}\\
&C_{2,i} = D_{6,i}(1-2D_4) +D_1^2t^{AA}_{Ii}\\
&C_{3,a} = D_{7,a}(1-2D_4) +D_1^2t^{aA}_{II}\\
&C_{4,ai} = 2D_{5,ai}D_1\\
&C_{5} = D_4(1-D_4) - D_1^2t^{AA}_{II}\\
&C_{6,i} = 2D_{6,i}D_1\\
&C_{7,a} = 2D_{7,a}D_1\\
&C_{8} = -D_1(1-2D_4)\\
&C_{9} = -D_1^2
\end{align}
\vspace{1cm}

and
\begin{align}
   &D_{4} = (D_2 - D_1 t^{A}_{I})\\
\begin{split}
  &D_{5,ai} = \frac{1}{2}\Big(-D_1(\delta_{AI,ai} + u^{Aa}_{Ii} -t^{A}_{i}t^{a}_{I}) \\
   &\phantom{D_{5,ai} =\frac{1}{2}\Big(} + D_3 t^{A}_{i}\delta_{Aa} -D_2 t^{a}_{I}\delta_{iI}\Big)
 \end{split}\\
   &D_{6,i} = \frac{1}{2}(D_1 t^{A}_{i} - D_2\delta_{Ii})\\
   &D_{7,a} = \frac{1}{2}(D_1 t^{a}_{I} + D_3\delta_{Aa}),
\end{align}
and where $u_{ij}^{ab} = 2t_{ij}^{ab} - t_{ji}^{ab}$. 

Finally, for $\bra{\mu_2}\simP$ we can show that the explicit expression becomes
\begin{widetext}
\begin{align}
\begin{split}
 \brad\simP = (D_4&-D_4^2- D_1^2t^{AA}_{II})\brad - (1-2D_4)D_1\brad\Eb{IA} -D_1^2\brad\Ea{IA}\Eb{IA}\\
+ P^{ab}_{ij}\Bigg[&
\Big\{(1-2D_4)G_{5,aibj} + F_{1,aibj} + F_{6,aibj} 
-G_{2,ia}D_{5,bj} -G_{8,bi}D_{5,aj} +
D_1(G_{1ib}t^{Aa}_{Ij} + G_{6,aij}t^{Ab}_{II} + G_{7,aib}t^{AA}_{Ij})\Big\}\braR\\
&+\Big\{(1-2D_4)G_{2,ia} + F_{3,ai} + F_{9,ai}\Big\}\brabj
+\Big\{(1-2D_4)G_{8,ib}\Big\}\bra{\bar{^{a}_j}}
+\Big\{(1-2D_4)G_{1,bi} + F_{2,ib} + F_{10,bi}\Big\}\braR\Eb{ja}\\
&+\Big\{(1-2D_4)G_{6,aij} + F_{4,jai} -D_{5,ai}G_{3,j} + F_{8,aij} -G_{2,ia}D_{6,j} -G_{8aj}D_{6,i}\Big\}\braR\Eb{Ib}\\
&+\Big\{(1-2D_4)G_{7,abi} + F_{5,bai} -D_{5,ai}G_{4,b} + F_{7,aib} -G_{2,ia}D_{7,b} -G_{8,bi}D_{7,a}\Big\}\braR\Eb{jA}\\
&+\Big\{2D_1G_{5,aibj}\Big\}\braR\Eb{IA}
+\Big\{2D_1G_{1,bi}\Big\}\braR\Eb{IA}\Ea{ja}
+\Big\{2D_1G_{6,aij}\Big\}\braR\Eb{IA}\Ea{Ib}\\
&+\Big\{2D_1G_{7,aib}\Big\}\braR\Eb{IA}\Ea{jA}
+\Big\{2D_1G_{2,ai}\Big\}\brabj\Ea{IA}
+\Big\{2D_1G_{8,bi}\Big\}\bra{\bar{^{a}_j}}\Ea{IA}\\
&+\Big\{(1-2D_4)G_{3,i} + 4D_1^2t^{AA}_{Ii}\Big\}\brabj\Eb{Ia}
+\Big\{(1-2D_4)G_{4,a} + 4D_1^2t^{aA}_{II}\Big\}\brabj\Eb{iA}\\
&+\Big\{-D_1^2t^{AA}_{ij} -D_{6,j}G_{3,i}\Big\}\braR\Eb{Ib}\Ea{Ia}
+\Big\{-D_1^2t^{ab}_{II} -D_{7,b}G_{4,a}\Big\}\braR\Eb{jA}\Ea{iA}\\
&+\Big\{-2D_1^2t^{Ab}_{iI} -D_{7,b}G_{3,i} -D_{6,i}G_{4,b}\Big\}\braR\Eb{jA}\Ea{Ia}
+\Big\{2D_1G_{3,i}\Big\}\brabj\Ea{IA}\Eb{Ia}
+\Big\{2D_1G_{4,a}\Big\}\brabj\Ea{IA}\Eb{iA}\revS{\Bigg]}
\end{split}\label{eq:doubles}
\end{align}
\end{widetext}
where we have defined
\begin{align}
\begin{split}
    &G_{1,bi} = D_2 t^{b}_{I}\delta_{Ii} - D_3 t^{A}_{i}\delta_{Ab} \\
    &\phantom{G_{1,bi} =}-D_1(t^{bA}_{Ii}+t^{A}_{i}t^{b}_{I}- \delta_{AI,bi})
\end{split}\\
\begin{split}
    &G_{2, ai} = -2D_1(\delta_{AI,ai} + u^{Aa}_{Ii} - t^{A}_{i}t^{a}_{I})\\
    &\phantom{G_{2,ai} =}+2D_3 t^{A}_{i}\delta_{Aa} - 2D_2 t^{a}_{I}\delta_{Ii}
\end{split}\\
    &G_{3,i} = 2(t^{A}_{i}D_1 - D_2 \delta_{Ii})\\
    &G_{4,a} = 2(t^{a}_{I}D_1 + D_3 \delta_{Aa})
\end{align}
\begin{align}
\begin{split}
    &G_{5,aibj} = D_3 u^{Aa}_{ji}\delta_{Ab} - D_2 u^{ab}_{iI}\delta_{Ij}\\
    &\phantom{G_{5,aibj} =}+D_1(u^{Aa}_{ji}t^{b}_{I} + u^{ab}_{iI}t^{A}_{j})
\end{split}\\
    &G_{6,aij}= D_1 u^{Aa}_{ji}\\
    &G_{7,abi}= D_1 u^{ab}_{iI}\\
    &G_{8,bi} = 2D_1t^{Ab}_{Ii}
\end{align}
and
\begin{align}
\begin{split}
    &F_{1, aibj} = D_1^2(u^{Aa}_{ji}t^{bA}_{II} + u_{Ii}^{ba}t^{AA}_{jI} \\
    &\phantom{F_{1, aibj} =  -D_1^2(}- t^{AA}_{ji}t^{ba}_{II} - t^{Aa}_{jI}t^{Ab}_{iI})
\end{split}\\
\begin{split}
    &F_{2,ib} = -D_1((-D_2 t^{Ab}_{II}\delta_{Ii} + D_3 t^{AA}_{Ii}\delta_{Ab})  \\
    &\phantom{F_{2,ib} = -D_1(}+D_1(t^{AA}_{iI}t^{b}_{I} + t^{bA}_{II}t^{A}_i))
\end{split}
\end{align}
\begin{align}
\begin{split}
    &F_{3,ai} = -2D_1((-D_3 t^{AA}_{Ii}\delta_{Aa} + D_2 t^{Aa}_{II}\delta_{Ii}) \\
    &\phantom{F_{3,ai} = -2D_1((}-D_1(t^{AA}_{iI}t^{a}_{I} + t^{aA}_{II}t^{A}_{i}))
\end{split}\\
\begin{split}
    &F_{4,ibj} = -D_1((-D_2 t^{Ab}_{iI}\delta_{Ij} + D_3 t^{AA}_{ij} \delta_{bA}) \\
    &\phantom{F_{4,ibj} = -D_1((}+D_1(t^{AA}_{ji}t^{b}_{I} + t^{bA}_{Ii}t^{A}_{j}))
\end{split}\\
\begin{split}
    &F_{5,abj} = -D_1((-D_2 t^{ab}_{II}\delta_{Ij} + D_3 t^{aA}_{Ij}\delta_{bA}   \\
    &\phantom{F_{5,abj} = -D_1((}+D_1(t^{Aa}_{jI}t^{b}_{I} + t^{ba}_{II}t^{A}_{j}))
\end{split}\\
\begin{split}
  &F_{6,aibj} = \frac{1}{2}\Big[-D_1(t^{A}_{j}t^{ab}_{II} + t^{b}_{I}t^{Aa}_{jI})G_{3,i}\\
    &\phantom{F_{6,aibj} = \frac{1}{2}\Big[}-D_1(t^{A}_{j}t^{bA}_{Ii} + t^{b}_{I}t^{AA}_{ji})G_{4,a}\\
    &\phantom{F_{6,aibj} = \frac{1}{2}\Big[}+D_2\delta_{Ij}(G_{3,i}t^{ba}_{II} + G_{4,a}t^{bA}_{Ii})\\
    &\phantom{F_{6,aibj} = \frac{1}{2}\Big[}-D_3\delta_{Ab}(G_{3,i}t^{Aa}_{jI} + G_{4,a}t^{AA}_{ji})\Big]
\end{split}\\
&F_{7,aib} = -\frac{1}{2}D_1(t^{ba}_{II}G_{3,i} + t^{bA}_{Ii}G_{4,a} + t^{Ab}_{Ii}G_{4,a})\\
&F_{8,aij} = -\frac{1}{2}D_1(t^{AA}_{ji}G_{4,a} + t^{Aa}_{jI}G_{3,i} + t^{Aa}_{Ij}G_{3,i}) \\
&F_{9,ai} =  D_1(t^{Aa}_{II}G_{3,i} + t^{AA}_{Ii}G_{4,a})\\
&F_{10,bi} = -\frac{1}{2}D_1(t^{AA}_{Ii}G_{4,b} + t^{Ab}_{II}G_{3,i}).
\end{align}
From Eqs. \eqref{eq:reference} -- \eqref{eq:doubles}, we see that  $\braR\simP$, $\bra{\mu_1}\simP$, and $\bra{\mu_2}\simP$ have contributions from the reference, singly and doubly excited determinants, and that 
$\bra{\mu_1}\simP$ and $\bra{\mu_2}\simP$ have contributions from triply excited determinants where two indices equal $I$ and $A$, and finally that there is a contribution from the quadruply excited determinant $\brad\Ea{IA}\Eb{IA}$ to $\bra{\mu_2}\simP$.
Contributions from the reference, and singly and doubly excited determinants, can be obtained from a standard closed-shell CCSD code. The new terms that must be implemented are the triply and quadruply excited determinants. Because of the restricted indices (to $A$ and $I$) in these terms, the correct scaling is $\mathcal{O}(N^6)$. However, in our naive implementation the scaling is the same as in CCSDTQ---$\mathcal{O}(N^{10})$, as we calculate the contribution from the triply excited determinants, and, more importantly, the quadruply excited determinant, without exploiting the index restrictions. 

\bibliography{main}

\end{document}